\documentclass[reprint,nofootinbib,showpacs,USenglish,prd]{revtex4-1}
\pdfoutput=1

\usepackage{preprintcover}
\PreprintCoverPaperTitle{\AtlasTitleText}
\PreprintIdNumber{CERN-PH-EP-2014-284}
\PreprintCoverAbstract{\AtlasAbstractText}
\PreprintJournalName{Physical Review D}

\usepackage{atlaspackage}
\usepackage{atlasphysics}
\usepackage{footnote}

\usepackage[coverpage=false]{atlascover}

\newcommand{\hath}[1]{\hat{\hat{#1}}}

\newcommand{\ttg}{\ensuremath{\ttbar\gamma}}
\newcommand{\pp}{\ensuremath{pp}}

\def \lumi {\cal L}
\newcommand{\Zgstar}{Z^0\kern -0.25em/\kern -0.15em\gamma^*}

\newcommand{\ZeejetsPar}{\ensuremath{Z(\rightarrow e^+e^-)+\ge 4}-jets}
\newcommand{\Wg}{W\gamma}
\newcommand{\Wgjets}{W\gamma \kern -0.05em + \kern -0.05em {\rm jets}}
\newcommand{\Zjets}{Z \kern -0.05em + \kern -0.05em {\rm jets}}
\newcommand{\Zgjets}{Z\gamma \kern -0.05em + \kern -0.05em {\rm jets}}
\newcommand{\Zee}{Z\rightarrow e^+e^-}
\newcommand{\ZeePar}{Z(\rightarrow e^+e^-)}

\newcommand{\Zll}{Z\rightarrow \ell^+\ell^-}

\newcommand{\Zg}{Z^0\kern -0.25em/\kern -0.15em\gamma^*}

\newcommand{\ptcone}{\mbox{$\pt^{\mathrm{iso}}$}}
\newcommand{\meg}{\mbox{$m_{e\gamma}$}}
\newcommand{\vect}[1]{\vec{#1}}
\newcommand{\mtw}{\ensuremath{m_{\mathrm{T}}(W)}}

\def\sttgfid{\sigma_{\ttg}^{\text{fid}}}
\def\sttgfidbr{\sttgfid \times {\rm BR}}
\def\afake{\alpha_\text{fake}}
\newcommand{\hideit}[1]{{ }}

\newcommand{\runilumi}{{\rm 4.59~fb^{-1}}}
\newcommand{\runilumierr}{{\rm 4.59 \pm 0.08~fb^{-1}}}
\newcommand{\cmu}{34.3 \pm 1.0\ (\text{stat.})\%}
\newcommand{\cel}{ 17.8  \pm 0.5\ (\text{stat.})\%}

\newcommand{\jimmyv}{{\tt JIMMY~4.31}}

\newcommand{\photosv}{{\tt PHOTOS~v2.15}}
\newcommand{\whizard}{{\tt WHIZARD}}
\newcommand{\whizardv}{{\tt WHIZARD~v1.93}}
\newcommand{\geant}{{\tt GEANT4}}
\newcommand{\herwig}{{\tt HERWIG}}
\newcommand{\herwigv}{{\tt HERWIG~v6.520}}
\newcommand{\mcatnlo}{{\tt MC@NLO}}
\newcommand{\mcatnlov}{{\tt MC@NLO~v3.1}}

\newcommand{\pythia}{{\tt PYTHIA}}

\newcommand{\pythiav}{{\tt PYTHIA~v6.425}}
\newcommand{\acermc}{{\tt AcerMC}}
\newcommand{\acermcv}{{\tt AcerMC~v3.8}}
\newcommand{\alpgen}{{\tt ALPGEN}}
\newcommand{\alpgenv}{{\tt ALPGEN~v2.13}}
\newcommand{\madgraph}{{\tt MadGraph}}
\newcommand{\madgraphv}{{\tt MadGraph~v5.1.5.12}}
\newcommand{\deltarmadgraph}{\Delta R}
\newcommand{\dr}{\Delta R}
\newcommand{\pmasym}[2]{^{+#1}_{-#2}}

\newcommand{\sherpav}{{\tt SHERPA~v1.4.0}}

\newcommand{\topmcv}{{\tt Top++~v2.0}}
\newcommand{\perugia}{{\tt PERUGIA}}

\def\thisttbar{The contribution from multijet production and its
  uncertainties are estimated using a data-based technique (see
  Sec.~\ref{s:qcd}). Other contributions are estimated using Monte
  Carlo simulations. The uncertainty band includes statistical and
  systematic uncertainties. The systematic uncertainties include
  those on lepton, jet, $\met$, and  $b$-tagging modeling, as well as systematic
  uncertainties on the multijet background estimate. 
}

\def\thisttbarphoton{The contribution from multijet+$\gamma$
  production and its uncertainties are estimated using a data-based
  technique (see Sec.~\ref{s:qcd}). The remaining contributions are
  estimated using Monte Carlo simulations. Other backgrounds (labeled
  as `Other bck.') include contributions from \Zjets, dibosons and
  single-top-quark production. }

\def\thisttbarphotonsystematics{The uncertainty band includes
  statistical and systematic uncertainties. The systematic
  uncertainties include those on photon, lepton, jet, $\met$, and
  $b$-tagging modeling, as well as systematic uncertainties for the
  multijet background estimate. }

\def\thiswhizardvsmcatnlo{The contribution from $\ttbar$ production with
  prompt photons (labeled as `$\ttg$') is estimated using the
  \whizard\ $\ttg$ Monte Carlo simulation. The contribution from $\ttbar$
  events with electrons and hadrons misidentified as prompt photons is
  obtained using inclusive $\ttbar$ Monte Carlo simulation. }

\def\thisoverflow{The last bin contains any overflow. }

\def\thisdottedlinerules{The horizontal dotted line corresponds to a
  value of $-\log\left[\lambda_s( \ptcone\,|\, \sigma^\text{fid}_{\ttg})\right]
  =0.5$. Intersections of this line with the solid (dashed) curve give the
  $\pm 1 \sigma$ total (statistical only) uncertainty interval to the
  measured fiducial $\ttg$ cross section. }

\newcommand{\sigCombel}{52}
\newcommand{\sigCombelErr}{14}
\newcommand{\sigCombmu}{100}
\newcommand{\sigCombmuErr}{28}
\newcommand{\sigComTot}{152}
\newcommand{\sigComErr}{31}
\newcommand{\bckCombTot}{199}
\newcommand{\bckCombTotErr}{47}
\newcommand{\bkgttgel}{79~\pm~ 26} 
\newcommand{\bkgttgmu}{120~\pm~39}
\newcommand{\hadFakesCombel}{38} 
\newcommand{\hadFakesCombelErr}{26} 
\newcommand{\hadFakesCombmu}{55} 
\newcommand{\hadFakesCombmuErr}{38} 
\newcommand{\hadFakesComb}{93} 
\newcommand{\hadFakesCombErr}{46}
\newcommand{\promptPhotonsComb}{106}
\newcommand{\promptPhotonsCombErr}{10}
\newcommand{\promptPhotonsCombEl}{41} 
\newcommand{\promptPhotonsCombElErr}{5} 
\newcommand{\promptPhotonsCombMu}{65}
\newcommand{\promptPhotonsCombMuErr}{9}
\newcommand{\mysignificance}{5.3}

\newcommand{\ttgel}{140}
\newcommand{\ttgmu}{222}
\newcommand{\xSecMin}{63}

\newcommand{\Systerrargh} {^{+17}_{-13}}

\newcommand{\totStat}{\pm 8}
\newcommand{\totSysUp}{+17}
\newcommand{\totSysDown}{-13}
\newcommand{\totLumi}{1}
\newcommand{\xsectot}{\xSecMin ^{+19}_{-16} \, \, \mathrm{fb}}
\newcommand{\xsecexp} {\sttgfidbr=  \xSecMin \, \totStat \mathrm{(stat.)}\,^{\totSysUp}_{\totSysDown} \, \mathrm{(syst.)} \pm \totLumi \,\mathrm{(lumi.) } \, \mathrm{fb}}
\newcommand{\xsecexpfb} {\xsecexp}
\newcommand{\xsecexpel}  {\sttgfidbr=76 ^{+16}_{-15}\mathrm{(stat.)}\,^{+22}_{-17} \, \mathrm{(syst.)}\pm 1\mathrm{(lumi.)} \, \mathrm{fb}}
\newcommand{\xsecexpelfb}{\xsecexpel}
\newcommand{\xsecexpmu}  {\sttgfidbr=55 ^{+10}_{-9} \mathrm{(stat.)}\,^{+14}_{-11} \, \mathrm{(syst.)}\pm 1 \mathrm{(lumi.)} \, \mathrm{fb}}
\newcommand{\xsecexpmufb}{\xsecexpmu}

\newcommand{\xsectheorfb}{48 \pm10\,\, \mathrm{fb}}
\newcommand{\xsectheorfbMG}{47 \pm10\, \mathrm{fb}}
\newcommand{\pvalue}{p_{0}^{\rm{obs}}= 5.73  \times 10^{-8}}

\newcommand{\LLratioObsVal}{14.1}

\def\paperTitle{Observation of top-quark pair production in association with a 
photon and measurement of the $\boldsymbol{\ttg}$ production cross 
section in {\bf{\em pp}} collisions at $\bf\sqrt{s}=7~\TeV$ using the ATLAS detector }

\def\paperAbstract{%
  A search is performed for top-quark pairs
  ($\ttbar$) produced together with a photon ($\gamma$) with
  transverse energy greater than 20~\gev\ using a sample of $\ttbar$ candidate
  events in final states with jets, missing transverse momentum, and one
isolated electron or muon. The  dataset used corresponds to an integrated luminosity of $\runilumi$
  of proton--proton collisions at a center-of-mass energy of $7~\TeV$
  recorded by the ATLAS detector at the CERN Large Hadron Collider. In
  total $\ttgel$ and $\ttgmu$ $\ttg$ candidate events are observed in
  the electron and muon channels, to be compared to the expectation of
  $\bkgttgel$ and $\bkgttgmu$ non-$\ttg$ background events
  respectively. The production of $\ttg$ events is observed with a
  significance of $\mysignificance$ standard deviations away from the null
  hypothesis.  The $\ttg$ production cross section times the 
branching ratio (BR) of the single-lepton decay channel
  is measured in a fiducial kinematic region within the ATLAS
  acceptance. The measured value is \mbox{$\xsecexpfb$} per lepton
  flavor, in good agreement with the leading-order theoretical
  calculation normalized to the next-to-leading-order theoretical
  prediction of $\xsectheorfb$. }
  
\AtlasTitle{\paperTitle}

\AtlasAbstract{%
  \paperAbstract
}

\begin{document}

\title{\paperTitle}
\author{The ATLAS Collaboration}

\begin{abstract}
\paperAbstract
\end{abstract}

\maketitle

\section{Introduction}
\label{s:introduction}

Due to its large mass, the top-quark is speculated to play a special
role in electroweak symmetry breaking (EWSB). 
New physics connected
with EWSB can manifest itself in top-quark observables. For instance, top-quark 
couplings can be modified significantly in some extensions of
the Standard Model (SM). 
A measured yield of top-quark pair production in association with a photon ($\ttbar\gamma$)
can constrain models of new physics, for example those with composite top-quarks~\cite{Lillie:2007hd},
or with excited top-quark production, followed by the radiative decay $t^*\rightarrow t\gamma$. 
The $\ttbar \gamma$ coupling may be determined via an
analysis of direct production of top-quark pairs in
association with a photon, evidence of which was first
reported~\cite{CDFttg3} by the CDF collaboration.

In this paper, observation of top-quark pair production in
association with a photon in proton--proton ($pp$) collisions at
a center-of-mass energy of 
$\sqrt{s} = 7~\tev$ is presented using the full 2011 ATLAS data sample,
which corresponds to an integrated luminosity of $\runilumi$. This
analysis is performed on $\ttbar$ candidate events in the lepton plus jets
final state. 
The $\ttg$ candidates are the subset of $\ttbar$ candidate events with
an additional photon. 
The measurement of the $\ttg$ production cross section times the
branching ratio (BR) of the single-lepton decay channel ($\ell\nu_\ell
q\bar{q'}b\bar{b}\gamma$, where $\ell$ is an electron or muon) is
reported in a fiducial kinematic region within the ATLAS acceptance.

The paper is organized as follows. The ATLAS detector is briefly
described in Sec.~\ref{s:detector}. The data and Monte Carlo
simulation samples used in the analysis are described in
Sec.~\ref{s:samples}, followed by a description of the event
selection in Sec.~\ref{s:selection}. The definition of the fiducial
phase space used in the measurement is presented in
Sec.~\ref{s:fiducial}. The cross section is extracted from a
template-based profile likelihood fit using the photon track-isolation
distribution as the discriminating variable. Section~\ref{s:strategy}
details the overall strategy of the measurement, and describes how
prompt-photon and background templates are obtained. Background
estimates are discussed in Sec.~\ref{s:backgrounds}. An overview of
the systematic uncertainties in the measurement is presented in
Sec.~\ref{s:systematics}. Section~\ref{s:results} presents the results
of the measurement, followed by conclusions in
Sec.~\ref{s:summary}.

\section{Detector}
\label{s:detector}

A detailed description of the ATLAS detector can be found in
Ref.~\cite{Aad:2008zzm}. The innermost part of the detector is a
tracking system that is immersed in a 2~T axial magnetic field and
measures the momentum of charged particles within a pseudorapidity
range of ${|\eta|<2.5}$~\footnotemark[1]. 
The inner detector (ID) comprises silicon pixel and
microstrip detectors, and a transition radiation tracker. The
calorimeter system 
is composed of sampling electromagnetic and hadronic compartments with
either liquid argon or scintillator tiles as the active media. It
resides outside the ID, covering $|\eta|<4.9$. 
The outermost system is 
a muon spectrometer that is used to identify
and measure the momentum of muons in a toroidal magnetic field in the
region $|\eta|<2.7$, with detectors used for triggering within
$|\eta|<2.4$. A three-level trigger system selects the potentially
interesting events that are recorded for offline analysis.

\footnotetext[1]{ATLAS uses a right-handed coordinate system with its origin at the nominal interaction point (IP) in the center of the detector and the $z$-axis along the beam pipe. The $x$-axis points from the IP to the center of the LHC ring, and the $y$-axis points upward. Cylindrical coordinates $(r,\phi)$ are used in the transverse plane, $\phi$ being the azimuthal angle around the beam pipe. The pseudorapidity is defined in terms of the polar angle $\theta$ as $\eta=-\ln\tan(\theta/2)$. Transverse momentum and energy are defined as $\pt = p\sin\theta$ and $\ET = E\sin\theta$, respectively.}

\clearpage

\section{Data and Monte Carlo samples}
\label{s:samples}

Data recorded by the ATLAS detector in 2011 in $\pp$ collisions at 
$\sqrt{s}=7~\tev$ are considered for analysis. 
Requirements are imposed on the collected data to ensure the quality of the beam conditions
 and detector performance.
The total integrated luminosity of the analyzed data sample is
$\mathcal{L}=\runilumierr$~\cite{Aad:2013ucp}.

Monte Carlo simulation samples are used to study signal and background
processes, using the ATLAS detector simulation~\cite{:2010wqa} based
on the \geant\ program~\cite{AGO-0301}. To simulate effects of
multiple $pp$ interactions per bunch crossing (`pile-up'), all Monte
Carlo events are overlaid with additional inelastic events generated
with 
\pythia~\cite{Sjostrand:2006za} using the AMBT1 set of parameters 
(tune)~\cite{ATLAS-CONF-2010-031}.  
The events are then reweighted to
match the distribution of the 
mean number of interactions per
bunch crossing in the data. Simulated events are reconstructed in the
same manner as the data.

Signal \mbox{$\ttg$ events} with single-lepton ($\ell\nu_\ell
q\bar{q'}b\bar{b}\gamma$, $\ell\equiv$~$e,\,\mu,\,\tau$) or dilepton
($\ell\nu_\ell \ell'\nu_{\ell'}b\bar{b}\gamma$, $\ell/\ell'\equiv$~$e,\,\mu,\,\tau$)
final states are simulated with two independent
leading-order (LO) matrix element~(ME) Monte Carlo
generators, \mbox{\whizardv~\cite{Whizard,Omega}} and
  \mbox{\madgraphv~\cite{Maltoni:2002qb}}, both using the
  CTEQ6L1~\cite{cteq6} LO parton distribution function (PDF)
set. Both calculations take into account interference effects between
radiative top-quark production and decay processes. Details on the
generator-level settings of the two signal Monte Carlo samples are
available in Sec.~\ref{s:ttgammalo}. 
In the $\ttg$ and inclusive
$\ttbar$ samples the top-quark mass is set to $m_t = 172.5~\gev$.

The \whizard\ sample is interfaced to \herwigv~\cite{COR-0001} for
the parton showering and \jimmyv~\cite{JButterworth:1996zw} is used
for the underlying-event simulation. The AUET2
tune~\cite{ATLAS:2011gmi} is used. 
The \madgraph\ sample is interfaced 
to either the \pythiav\ parton shower using the
\perugia\ 2011 C tune~\cite{Skands:2010ak}, or with
\herwigv\ and \jimmyv\ for
the parton showering and the underlying-event simulations
respectively. 
\pythia\ QED final-state radiation (FSR) from charged hadrons and leptons
is switched off and instead \photosv~\cite{Photos} is used.

To compare with the experimental measurement, the LO calculations of
\whizard\ and \madgraph\ are normalized to the next-to-leading-order
(NLO) cross section, obtained for $\sqrt{s}=7~\TeV$ at the
renormalization and factorization scales of $m_t$. The
NLO QCD calculation of top-quark pair production in
association with a hard photon is detailed in Sec.~\ref{s:ttgammanlo}.
The systematic uncertainty on the NLO cross section is obtained by
simultaneous renormalization and factorization scale variations by a
factor of two ($m_t/2$ and $2 m_t$) around the central value ($m_t$), and is calculated to
be 20\%~\cite{Melnikov:private}. 
The NLO/LO correction ($K$-factor) calculation is performed in a phase-space region close to
the one defined by the analysis kinematic selection criteria (see Sec.~\ref{s:ttgammanlo} for details). The
dependence of the $K$-factor on the kinematic variables is small
compared to the scale uncertainty~\cite{Melnikov:private}. 

The effect of the variations of photon radiation settings in
\madgraph\ is studied using a sample generated with a minimum
separation in $\eta$--$\phi$ space between the photon and any other 
particle of $\deltarmadgraph > 0.05$~\footnotemark[2]\footnotetext[2]{$\Delta R = \sqrt{(\Delta\phi)^2 + (\Delta\eta)^2}$, where $\Delta\eta$ ($\Delta\phi$) is the separation in $\eta$ ($\phi$) between the objects in the $\eta$--$\phi$ space.} 
instead of $\deltarmadgraph > 0.2$ used in the default sample (see
Sec.~\ref{s:ttgammalo}). For this sample, \pythia\ QED FSR is switched
off and no additional photon radiation is produced by
\photosv. 
In addition to the default \madgraph+\pythia\ Monte Carlo sample
generated at the scale of $m_t$, samples at scales of $m_t/2$ and $2 m_t$ 
are produced to study the effect of scale variations. 

The simulated sample for inclusive $\ttbar$ production is generated
with \mcatnlov~\cite{Frixione:2002ik, Frixione:2003ei} (NLO ME $2\rightarrow 2$) interfaced to \herwigv\ for the
parton showering and fragmentation and to \jimmyv\ for underlying-event simulation, using the
CTEQ6.6~\cite{Nadolsky:2008zw} PDF set, with additional photon
radiation simulated with \photosv. This sample is used to validate distributions of kinematic variables 
in $\ttbar$ candidate events as described in Sec.~\ref{s:selection}.

Initial- and final-state QCD radiation (ISR/FSR) variations are
studied using inclusive $\ttbar$ samples generated with
\acermcv~\cite{Kersevan:2013ji} interfaced to \pythiav\ with the CTEQ6L1 PDF set.
In these samples the parameters that control the amount of ISR/FSR are set to values consistent with the
\perugia\ Hard/Soft tune in a range given by
current experimental data~\cite{ATLAS:2012al}. 
\acermcv\ $\ttbar$ samples interfaced to \pythiav\ are also used to
study variations of color reconnection using the \perugia\ 2011 C and
\perugia\ 2011 NO CR tunes~\cite{Skands:2010ak}. The underlying-event
variations are studied using \acermcv\ interfaced to \pythiav\
with two different underlying-event settings of the
AUET2B~\cite{ATLAS:2011zja} \pythia\ generator tune. In all these
\acermcv\ samples, photon radiation is simulated with
\photosv~\cite{Photos}. 
The inclusive $\ttbar$ signal samples are normalized to a
predicted Standard Model $\ttbar$ cross section of $\sigma_{t\bar{t}}=
177^{+10}_{-11}$~pb for a top-quark mass of $172.5~\GeV$, as obtained
at next-to-next-to-leading order (NNLO) in QCD including resummation of
next-to-next-to-leading-logarithmic (NNLL) soft gluon terms with
\topmcv~\cite{Cacciari:2011hy, Baernreuther:2012ws, Czakon:2012zr,
  Czakon:2012pz,Czakon:2013goa, Czakon:2011xx}. 

Background samples of $W$ and $Z$ bosons (including $W+b\bar{b}$ and
$Z+b\bar{b}$ processes) are generated with \alpgenv~\cite{MAN-0301}
interfaced to \herwigv, using the CTEQ6L1 PDF set. The
\alpgen\ matrix elements include diagrams with up to five additional
partons.
The MLM~\cite{MAN-0301} parton--jet matching
scheme is applied to avoid double counting of configurations generated
by both the parton shower and the LO matrix-element
calculation. 
In addition, overlap between heavy-flavor quarks that
originate from ME production and those that originate from the parton
shower is removed. 
Diboson ($WW$, $WZ$, and $ZZ$) production is
modeled using \herwigv\ and the MRST LO** PDF set~\cite{Martin:1998sq}. The $W\gamma$+jets 
and $Z\gamma$+jets (with up to three partons including $b \bar b$, $c\bar c$, $c$) processes are
generated with \sherpav~\cite{GLE-0901} and the CT10~\cite{Lai:2010vv}
NLO PDF set. Single-top-quark production is modeled using \acermc\ in the
$t$-channel and \mcatnlo~v3.41~\cite{Frixione:2008yi} for the $Wt$-
and~\cite{Frixione:2005vw} $s$-channels.  

Multijet samples with jet $\pt$ thresholds of 17, 35 and 70 GeV are
generated using \pythia\ v6.421 with the AUET2B~\cite{ATLAS:2011zja}
generator tune.

\section{Object and event selection}
\label{s:selection}

Events for the analysis are selected by requiring a high-$\pt$
single-electron or 
single-muon trigger~\cite{ATLASTrigPerformance2} for the electron and muon
channels respectively.
The $\pt$ threshold for the muon trigger is $18~\gev$, the thresholds
for the electron trigger are $20~\gev$ or $22~\gev$, depending on the
data-taking period due to changing LHC luminosity conditions. 
The event reconstruction makes use of kinematic variables such as
transverse momentum ($\pt$), energy in the transverse plane ($\et$)
and pseudorapidity ($\eta$) of photons, leptons ($e$ and $\mu$) and jets
($j$) as well as $b$-tagging information, and missing transverse
momentum ($\boldsymbol{\met}$).

The selected events are required to contain a reconstructed primary
vertex with at least five associated tracks, each with $\pt>0.4~\GeV$. The
primary vertex is chosen as the vertex with the highest $\sum{\pT^2}$
over all associated tracks.

Photons are required to have $\et>20~\GeV$ and $|\eta|<2.37$, excluding the transition region
between the barrel and endcap calorimeters at
1.37$<|\eta|<$1.52, and must satisfy
tight identification criteria~\cite{ATLAS:2012ar,ATLAS:2012ar2}. Specifically,
requirements on the electromagnetic shower shapes~\cite{Aad:2010sp}
are applied to suppress the background from hadron decays
(e.g. $\pi^0\rightarrow \gamma\gamma$ decay leads to two overlapping showers
as opposed to a single shower produced by a prompt photon).

Electrons~\cite{ATLASElectronPerformance} are reconstructed by
matching energy deposits in the electromagnetic calorimeter with
tracks in the ID, and are required to have $\et > 25~\gev$ and $|\eta|
< 2.47$, excluding the transition region between the barrel and endcap
calorimeters. 
Muons~\cite{Aad:2014zya} are reconstructed by 
matching tracks in the ID with tracks measured in the muon
spectrometer, and are required to have $\pt > 20~\gev$ and $|\eta| <
2.5$.

\begin{figure*}[!thbp]
\vskip-0.1in
\centering
\includegraphics[width=0.49\textwidth]{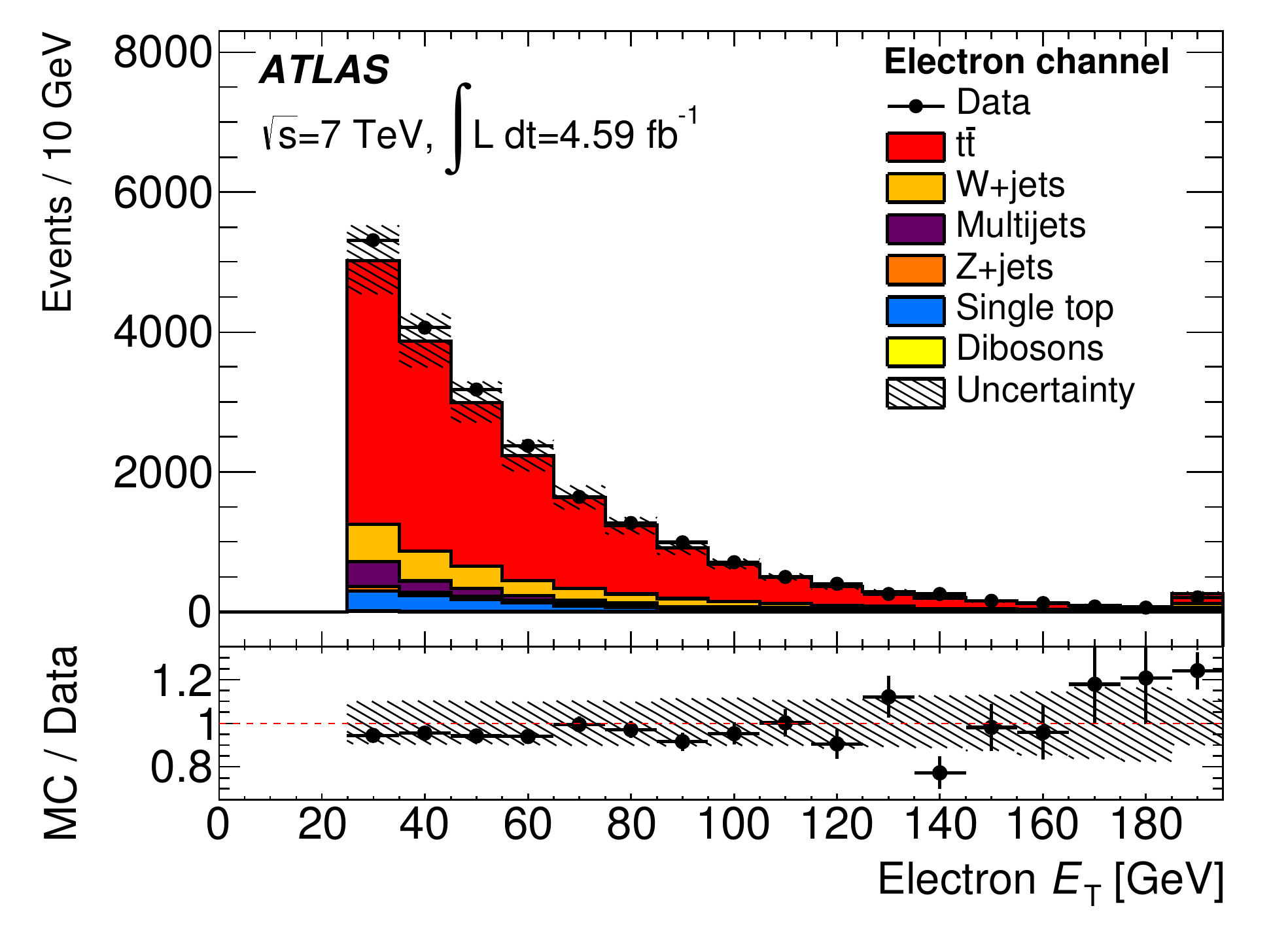}
\hspace{-0.1cm}
\includegraphics[width=0.49\textwidth]{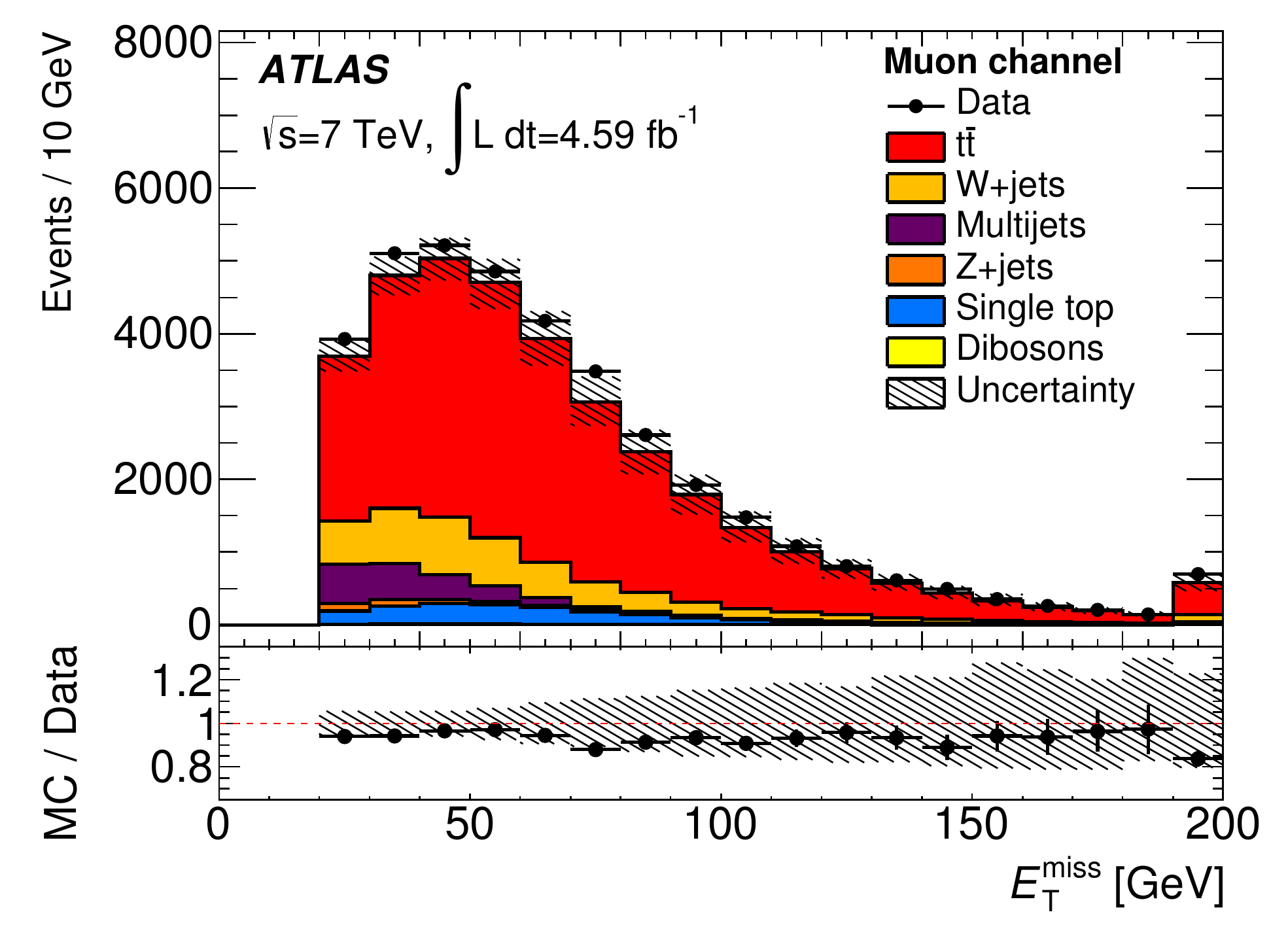}
\vspace{-0.2cm}
\caption{Comparison of distributions in data (points) versus
  expectation (stacked histograms) for the $\ttbar$
  selection (see text). The electron transverse energy ($\et$) in the
  electron channel is shown on the left, missing transverse momentum
  ($\met$) in the muon channel is shown on the right. \thisttbar
  \thisoverflow }
\label{f:ttbar_control}
\end{figure*}

Leptons are required to be isolated to reduce the number of lepton candidates 
that are misidentified hadrons or non-prompt leptons. To calculate the
isolation of electrons in the calorimeter, the $\et$ deposited in the
calorimeter in a cone of size $\Delta R=0.2$ around the electron is summed, and
the $\et$ due to the electron itself is subtracted. The scalar sum of 
$\pt$ of tracks with $\pT > 1~\gev$ originating from the primary
vertex in a cone of $\Delta R=0.3$ around the electron direction is
also measured, excluding the electron track. Selection requirements are 
parameterized as a function of the electron $\eta$ and $\et$ and
applied to these two isolation variables to ensure a constant
efficiency of the isolation criteria of 90\% (measured on $\Zee$ data) over the entire ($\eta$, $\et$) range. 
For muons, the transverse energy deposited in the calorimeter
in a cone of $\Delta R=0.2$ around the muon direction is required
to be less than $4~\gev$, after subtraction of the $\et$ due to the
muon itself. The scalar sum of the transverse momenta of tracks in a cone of
$\Delta R=0.3$ is required to be less than $2.5~\gev$ after
subtraction of the muon track $\pt$.
The efficiency of the muon isolation requirements
is of the order of 86\% in simulated $\ttbar$ events
in the muon+jets channel.

Jets~\cite{Aad:2011he} are reconstructed from topological
clusters~\cite{Cojocaru:2004jk,Lampl:2008zz} of energy deposits in the calorimeters
using the \mbox{anti-$k_t$}~\cite{antikt} algorithm with a distance
parameter $R~=~0.4$. Jets selected for the analysis are required to
have $\pt > 25~\gev$ and $|\eta| < 2.5$.
In order to reduce the background from jets originating from pile-up
interactions, 
the jet vertex fraction,
defined as the sum of the $\pT$ of tracks associated with the jet and
originating from the primary vertex divided by the sum of the $\pT$
from all tracks associated with the jet, is required to be greater
than 0.75. Since electrons and photons deposit energy in the calorimeter, they
can be reconstructed as jets. The jet closest to an
identified electron in $\eta$--$\phi$ space is rejected if $\Delta
R(e,j)<0.2$~\cite{Aad:2012qf}. Similarly, any jet within $\Delta
R(\gamma,j) = 0.1$ of an identified photon is discarded. 
To suppress muons from heavy-flavor hadron decays
inside jets, muon candidates within $\Delta R(\mu,j) < 0.4$ are
rejected~\cite{Aad:2012qf}.

\begin{figure*}[!thbp]
\vskip-0.1in
\centering
\includegraphics[width=0.49\textwidth]{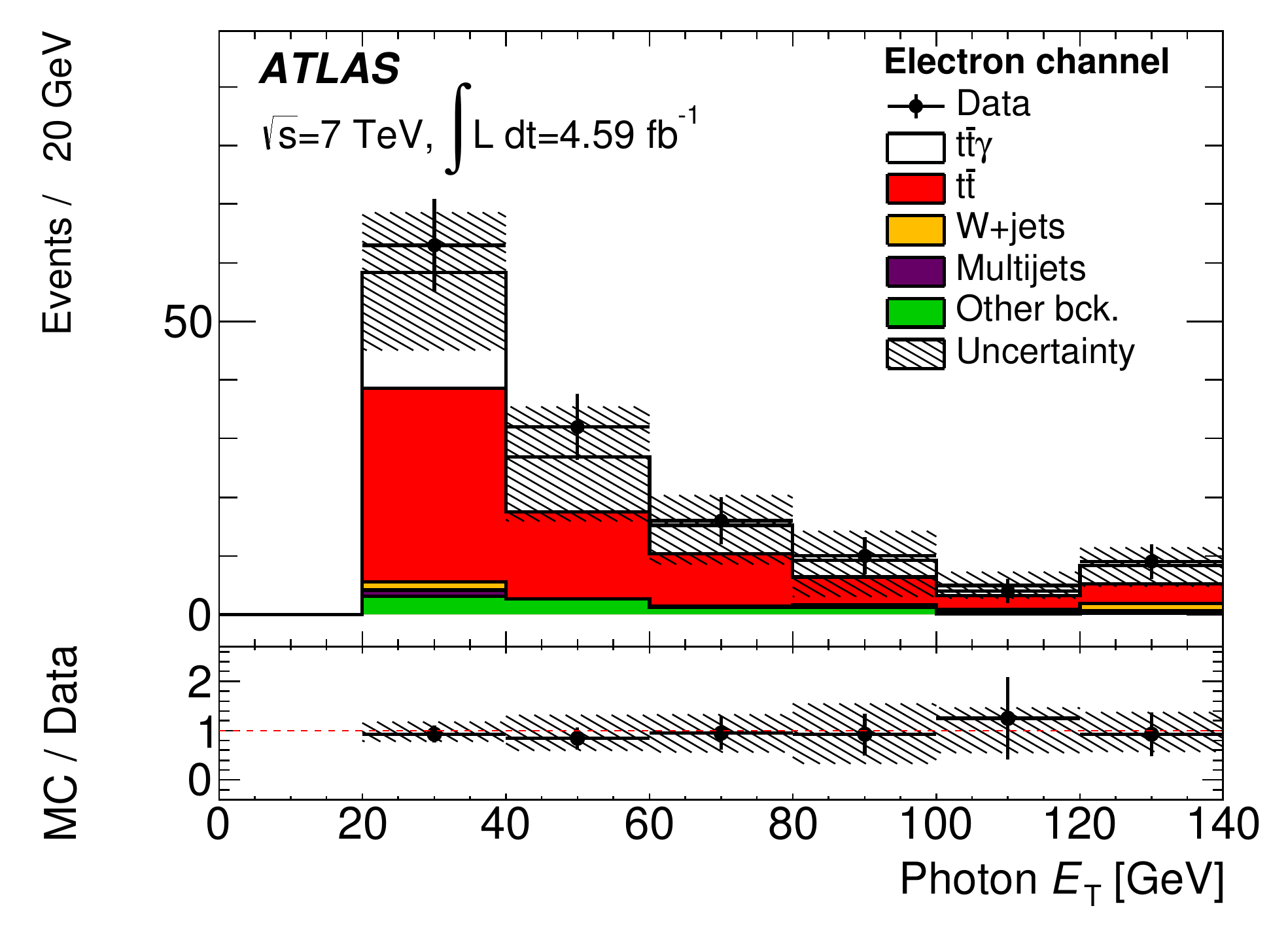}
\hspace{-0.1cm}
\includegraphics[width=0.49\textwidth]{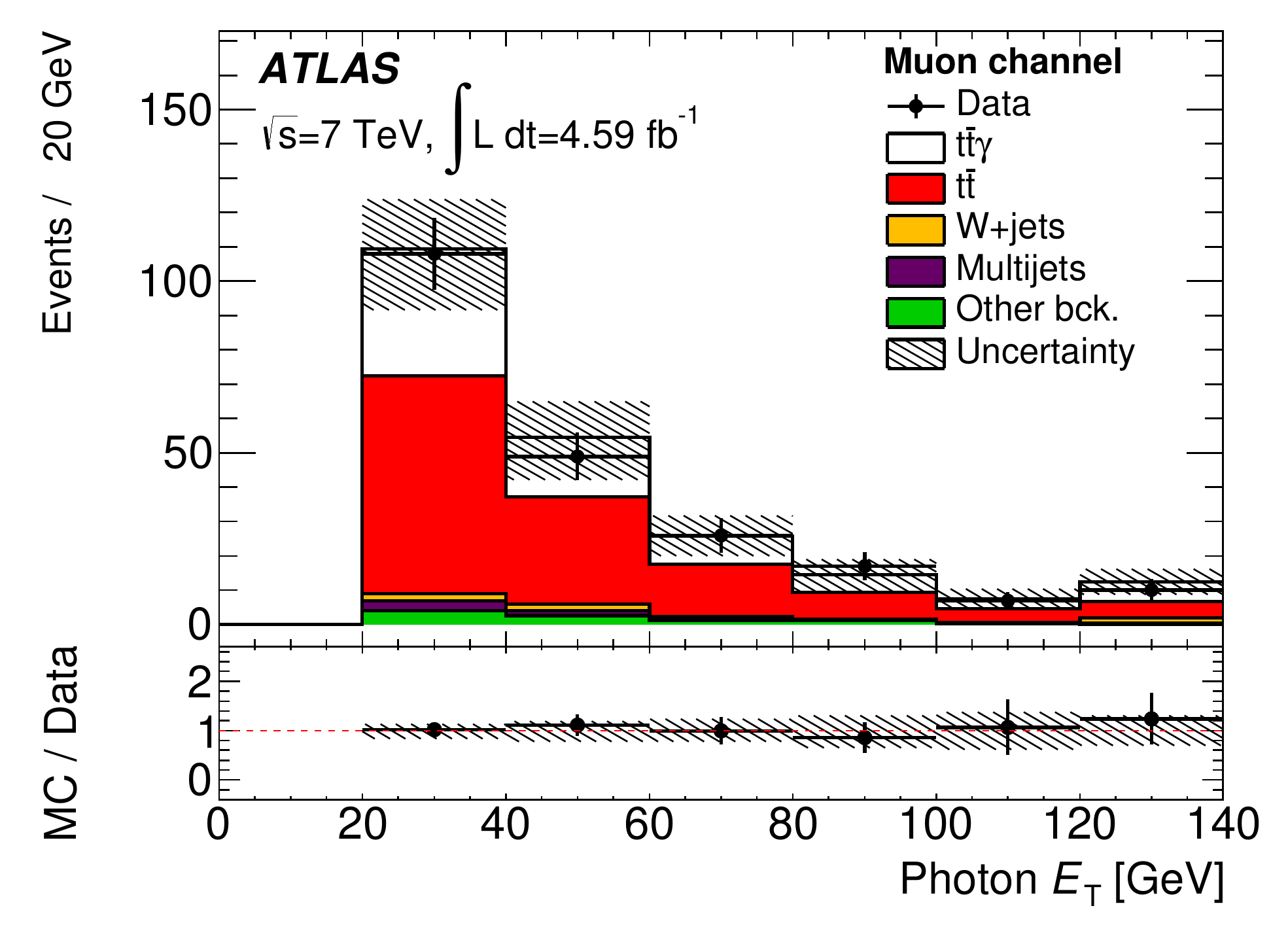}
\vspace{-0.2cm}
\caption{Distributions for the $\ttg$ selection (see
  text). The photon candidate transverse energy ($\et$) distribution in data (points) is
  compared to the expectation (stacked histograms)
  for the electron (left) and muon (right) channels. 
\thisttbarphoton 
\thiswhizardvsmcatnlo 
\thisttbarphotonsystematics
\thisoverflow}
\label{f:ttg_control}
\end{figure*}

Jets containing a $b$-hadron are identified with a $b$-tagging
algorithm~\cite{ATLAS-CONF-2012-040,ATLAS-CONF-2012-043,ATLAS-CONF-2011-089}
using impact parameter and vertex position
measurements from the inner detector as inputs to a neural network; 
$b$-tagged jets are required to satisfy a selection that is 70\%
efficient for $b$-quark jets in simulated $\ttbar$ events.
The misidentification rate of light-flavor partons ($u$, $d$, $s$-quark or gluon) is
in the range from 1\% to 3\%, depending on the jet \pt\ and $\eta$~\cite{ATLAS-CONF-2012-040}.

The transverse momentum of the neutrinos produced in the top-quark decay chains, measured 
as missing transverse momentum, is reconstructed from
the vector sum of the transverse momenta corresponding to all
calorimeter cell energies contained in topological
clusters~\cite{Aad:2011he} with \mbox{$|\eta|<4.9$}, projected onto
the transverse plane. Contributions to $\boldsymbol{\met}$ from the calorimeter
cells associated with physics objects (jets, leptons, photons) are
calibrated according to the physics object
calibration~\cite{Aad:2012re}.  The contribution to $\boldsymbol{\met}$ from the
$\pt$ of muons passing the selection requirements is included.
Calorimeter cells containing energy deposits above noise and not
associated with high-$\pT$ physics objects are also included.

Top-quark-pair candidate events are selected by requiring exactly one
lepton $\ell$ (where $\ell$ is an electron or muon) and at least four jets, of which at least one must be
$b$-tagged. \mbox{To reduce} the background from multijet processes, events
in the electron channel are required to have \mbox{$\MET>30~\GeV$}, where $\MET$ is the
magnitude of the missing transverse momentum $\boldsymbol{\met}$, and a $W$-boson
transverse mass $\mtw>35~\GeV$. This $W$-boson
transverse mass is defined as
\mbox{$\mtw=\sqrt{2 p_{\rm T}^\ell\times\MET(1-\cos\phi)}$}, where
$p_{\rm T}^\ell$ is the transverse momentum of the lepton 
and  $\phi$ is the azimuthal angle between the lepton direction and the missing transverse momentum vector. 
Similarly, events in the muon channel are required to have
$\MET>20~\GeV$ and $\mtw+\MET>60~\GeV$. 
Representative distributions of kinematic variables for this selection
are shown in Fig.~\ref{f:ttbar_control}. 

The analysis of $\ttg$ production is performed on the subset of
selected $\ttbar$ candidate events that contain at least one photon
candidate.  To suppress the contributions from photons radiated from
leptons, photon candidates with $\Delta R(\gamma,\ell) < 0.7$ are
discarded.  Events with a jet closer than $\Delta R(\gamma,j)=0.5$ in
$\eta$--$\phi$ space to any photon candidate are discarded, as those
photons have a reduced identification efficiency.  
In addition, to
suppress the contribution from $\ZeePar$+jets production with one
electron misidentified as a photon, the $e\gamma$ invariant mass
$\meg$ is required to be $|\meg-m_Z|>5~\GeV$, where $m_Z=91~\GeV$.  
This selection yields totals of $\ttgel$ and $\ttgmu$ events in data
in the electron and muon channels respectively. In Fig.~\ref{f:ttg_control} the photon
candidate $\et$ distributions for this selection are compared to
predictions for the electron and muon channels.

Corrections are applied to
simulated samples when calculating acceptances to account for
observed differences between predicted and observed trigger, photon and lepton
reconstruction and identification efficiencies and jet $b$-tagging efficiencies and mistag
rates, as well as smearing to match jet~\cite{Aad:2014bia}, photon and lepton
energy resolutions in
data~\cite{Aad:2014zya,ATLASElectronPerformance2010}.

\section{Definition of the fiducial phase space and cross section}
\label{s:fiducial}

To allow a comparison of the analysis results to theoretical
predictions, the cross section measurement is made within a fiducial
phase space defined in Monte Carlo simulation for $\ttbar\gamma$ decays 
in the single-lepton (electron or muon) final state. 
The particle-level prediction is constructed using final-state particles
with a lifetime longer than 10~ps.

Photons are required to originate from a non-hadron parent, which is
equivalent to the requirement for photons to originate from a
top-quark radiative decay or top-quark radiative production. Photons
are required to have $\pt > 20~\GeV$ and $|\eta| < 2.37$.

Leptons are defined as objects constructed from the four-momentum
combination of an electron (or muon) and all nearby photons in a cone
of size $\dr = 0.1$ in $\eta$--$\phi$ space centered on the
lepton. Leptons are required to originate from a non-hadron parent,
which is equivalent to the requirement for leptons to originate from
the $t \rightarrow Wb \rightarrow \ell \nu b$ decays. Leptons are
required to have $\pt > 20~\GeV$ and $|\eta| < 2.5$.

Decays of $\ttbar\gamma$ to the dilepton final states, as well as 
decays to the single-lepton final state with an electron or muon coming 
from a $\tau \rightarrow \ell\nu\nu_\tau$ decay are considered as non-fiducial
and are corrected for when calculating the cross section.

The anti-$k_t$~\cite{antikt} algorithm with a distance parameter
$R~=~0.4$ is used to form particle-level jets from all particles with
a lifetime longer than 10 ps, excluding muons and neutrinos. Particles
arising from pile-up interactions are not considered. Jets are required to 
have $p_{\rm T} > 25~\GeV$ and $|\eta| < 2.5$.

The removal of overlapping particles is performed in a manner consistent with 
the object and event selection described in Sec.~\ref{s:selection}. 
Any jet with $\dr(e, j) < 0.2$ or $\dr(\gamma,j) < 0.1$ is discarded; any muon with $\dr(\mu, j) < 0.4$ is discarded.
To suppress the contribution of photon radiation off a charged lepton, 
photons within $\dr(\gamma, \ell) < 0.7$ are discarded. 

For the determination of the $\ttg$ fiducial cross section
$\sttgfid$, exactly one lepton (electron or muon), at least one
photon, and four or more jets are required. At least one jet must
match a $b$-hadron. All simulated $b$-hadrons that are generated with
$\pt > 5~\gev$ are considered for the matching, and are required to
satisfy \mbox{$\Delta R$($b$-hadron, $j$) $<$ 0.4}. 
Events with $\dr(\gamma,j) < 0.5$ are discarded.

The fiducial cross section $\sttgfid$ is calculated as
\mbox{$\sttgfid =
  {N_s}/({\epsilon\cdot\mathcal{L}}$}). 
The number of
estimated $\ttg$ signal events is $N_s=N-N_b$, where $N$ and $N_b$ are
the number of observed $\ttg$ candidate events in data and the
estimated number of background events respectively. 
The efficiency $\epsilon$ is determined from $\ttg$ Monte Carlo
simulation as the ratio of the number of all events passing the $\ttg$
event selection to the total number of events generated in the
fiducial region. It is $\cel$ for the electron
channel and $\cmu$ for the muon channel. These numbers include
kinematic and geometric acceptance factors, as well as trigger,
reconstruction and identification efficiencies. 
The efficiency values also account for migrations into and out of the
fiducial phase space.

\section{Analysis strategy}
\label{s:strategy}

After the selection more than half of the events do not come 
from $\ttg$ production. The track-isolation distribution of the photon candidates is 
used to discriminate between signal photons and neutral hadron decays to 
final states with photons and hadrons misidentified as photons. For simplicity,  
neutral hadron decays to diphoton final states and hadrons misidentified as photons
are referred to hereafter as `hadron-fakes'.

The photon track-isolation variable \ptcone\ is defined as the scalar sum of
the transverse momenta of selected tracks in a cone of $\Delta R=0.2$
around the photon candidate. The track selection requires at least six
hits in the silicon pixel and microstrip detectors, including at least one
hit in the innermost layer in the pixel detector (except when the
track passes through one of the 2\% of pixel modules known to be not
operational), track \mbox{$\pt>1~\GeV$}, longitudinal impact parameter $|z_0|
< 1$~mm and transverse impact parameter $|d_0| < 1$~mm computed with
respect to the primary vertex. The tracks from photon conversions are
excluded.

Prompt-photon and background track-isolation templates are obtained
from data as described in Sec.~\ref{s:signal_template}
and~\ref{s:background_template}. The total number of events with
prompt photon-like objects (for simplicity referred to as `prompt
photons' unless noted otherwise) is extracted using a template-based
profile likelihood fit. The expected number of non-$\ttg$ events with
prompt photons, as summarized in Table~\ref{tab:backgrounds_summary},
is subtracted to calculate the fiducial cross section
$\sttgfid$. These steps are incorporated in a likelihood fit.

\subsection{Likelihood description}

A binned template fit maximizes the following extended Poisson
likelihood function, representing the Poisson probability to observe
$N$ data events given an expectation of $(N_s+N_b)$ events:

\begin{small}
\begin{equation*}
\begin{aligned}
L \left ( \mathbf{\ptcone }\,|\, N_s,N_{b} \right ) = \frac{(N_s+N_{b})^N }{N!}e^{ -(N_s+N_{b} )}  \times P(\mathbf{ \ptcone}\,|\, N_{s}+ N_{b} ) \times &\\
\prod_{i=1}^{n}  P(N_{b_i}\,|\,\hat{N}_{b_i}) \times P_{\rm{eff}} ( \mathbf{\varepsilon}\,|\,\hat{\varepsilon} ) \times
P_{\rm{lum}} ( \mathcal{L}\, |\, \hat{\mathcal{L} } ).
\end{aligned}
\end{equation*}
\end{small}
\noindent
For a given variable $x$, $P(x|\hat x)$ is the probability of $x$
given $\hat x$, where $\hat{x}$ denotes the unconditional maximum-likelihood
estimate of $x$. Therefore, $P_{\rm eff}( \varepsilon\,
|\,\hat{\varepsilon})$ describes the systematic uncertainties
affecting the combined signal efficiency and acceptance
$\varepsilon$; $P_{\rm{lum}}( \lumi\, |\,\hat{\lumi})$ describes the uncertainty
on the integrated luminosity $\lumi$; $P( N_{b_i}\, |\,\hat{N}_{b_i})$
describes the uncertainty on the $i$-th background component $b_i$; $n$ is the
number of background sources, $N_b = \sum_{i=1}^{n}{N_{b_{i}}}$.

The modeling of the signal and the different background sources can be
expressed as:

\begin{small}
\begin{equation*}
\begin{aligned}
P(\mathbf{ \ptcone}| N_s+ N_b ) &= f_{sb} F_s (\mathbf{ \ptcone})+(1-f_{sb}) \sum_{i=1}^{n} F_{b_{i}}( \mathbf{\ptcone}),
\end{aligned}
\end{equation*}
\end{small}
\noindent\ignorespacesafterend where $F_s (\ptcone)$ and $F_{b_{i}}(
\mathbf{\ptcone})$ are the probability density functions (pdf) for the
signal and the $i$-th background source respectively, with
$f_{sb}={N_s}/({N_s+N_b})$ being the signal purity. 
Each $F_{b_i}$ is normalized to the corresponding background expectation $N_{b_i}/N_b$.

Every systematic uncertainty is taken into account as an independent
nuisance parameter modeled by a Gaussian pdf $\mathcal{N}$.
In the likelihood, \mbox{$\vect{\varepsilon}=  \left(\varepsilon_{\text{electron channel}} ,\, \varepsilon_{\text{muon channel}}\right)$} and $N_{b_i}$ are
considered to be functions of the nuisance parameters $\vect{\theta}$ and
$\vect{\alpha_{i}}$ respectively.
Taking into account the probability distribution functions modeling
the different parameters, the expanded form of the likelihood used to
fit $N_{\text{bins}}$ of the $\ptcone$ distribution for an expectation
of $N_j$ events in each bin $j$ spanning the range $V_j$ reads:
\begin{multline}
L \left (\ptcone\,|\,  \sigma^\text{fid}_{\ttg},\,\vect{\varepsilon}(\vect{\theta}),\,\mathcal{L},\, N_{b_1}(\vect{\alpha_1}),\ldots,N_{b_n}(\vect{\alpha_n}) \right ) =  \\
 \underbrace{\prod_{c=1}^{N_{\text {channels}}} \left [ \prod_{j=1}^{N^{c}_{\text{bins}}}   \frac {\nu_{j}^{N_j}}{N_j!} \cdot e^{ \nu_{j}} \right ]}_\text{ Poisson  expectation} 
\,\,\times 
\underbrace{\prod_{l=1}^{N_{\text{bkg-syst}}} \mathcal{N}(\alpha_l|\hat{\alpha_{l}},\sigma_{\alpha_l} ) }_\text{background uncertainties} \\ 
\,\,\,\,\, \times   \!\!\!\!\!
\underbrace{\prod_{k=1}^{N_{\text{syst}}} \mathcal{N}(\theta_k|\hat{\theta_{k}},\sigma_{\theta_k} )}_\text{efficiency/acceptance uncertainties}  
\!\!\!  \times \,\,\,
\underbrace{\mathcal{N}(\mathcal{L}|\hat{\mathcal{L}}, \sigma_{\mathcal{L}} )}_\text{luminosity uncertainty},
\label{eq:LL}
\end{multline} 
\noindent where $\nu_j$ is defined as: 
\begin{multline}
\nu_{j} = \nu_{j}( \sigma^\text{fid}_{\ttg},\varepsilon_c(\vect{\theta}),\mathcal{L},N_{b_1}(\vect{\alpha_1}),\ldots,N_{b_n}(\vect{\alpha_n})  ) = \\ 
 \varepsilon_c(\vect{\theta})\mathcal{L}\sigma^\text{fid}_{\ttg}\int_{V_j} d\ptcone F^j_{s}(\ptcone | \sigma^\text{fid}_{\ttg})\, + \\ 
+ \sum_{i=1}^{n} N_{b_i} (\vect{\alpha}_i)\int_{V_j}d\ptcone F^j_{b_i}(\ptcone | N_{b_i} (\vect{\alpha}_i)),
\end{multline} 
\noindent with $c\!\equiv$\{electron channel, muon channel\}, and $i=1,\dots,N_{\text{bkg-syst}}$ and $k=1,\dots,N_{\text{syst}}$
denoting the systematic uncertainties on the background and the signal
efficiency/acceptance respectively.
The normal pdf, modeling the nuisance parameter $x$, is denoted by
$\mathcal{N}(x | \hat{x},\sigma_{x})$.
The $\ptcone$ binning is chosen to minimize the statistical uncertainty.

Finally, a profile likelihood ratio $\lambda_s$ is
built~\cite{RooFit,RooStats} by considering the cross section as the
parameter of interest and all other parameters to be nuisance parameters:
\begin{small}
\begin{equation*}
\lambda_s( \ptcone\,|\, \sigma^\text{fid}_{\ttg}) = \frac{ L  (\ptcone\,|\, \sigma^\text{fid}_{\ttg},\hath{\vect{\varepsilon}}(\vect{\theta}), \hath{\mathcal{L}}, \hath{N}_b(\vect{\alpha}))} 
{ L (\ptcone\,|\, \hat{ \sigma}^\text{fid}_{\ttg},\hat{\vect{\varepsilon}}(\vect{\theta}),\hat{\mathcal{L}} ,\,\hat{N}_{b}(\vect{\alpha}))}
\label{eq:LLratio}
\end{equation*}
\end{small}
\noindent
Here, for a given parameter $x$, $\hath{x}$ is the value of $x$ that
maximizes the likelihood function for a given $\sigma^\text{fid}_{\ttg}$. The
numerator thus depends on the conditional likelihood estimator of $x$,
and the denominator depends on the maximized (unconditional)
likelihood estimator.

\subsection{Prompt-photon template}
\label{s:signal_template}

The prompt-photon template models the \ptcone\ distribution of prompt
photons as well as electrons misidentified as photons, from $\ttg$ and
background processes. While the same template is used for prompt
photons and electrons misidentified as photons, the possible
differences are covered by alternative templates used to estimate the
systematic uncertainties as discussed below.

\begin{figure}[!bhtp]
\centering	
\includegraphics[width=0.5\textwidth]{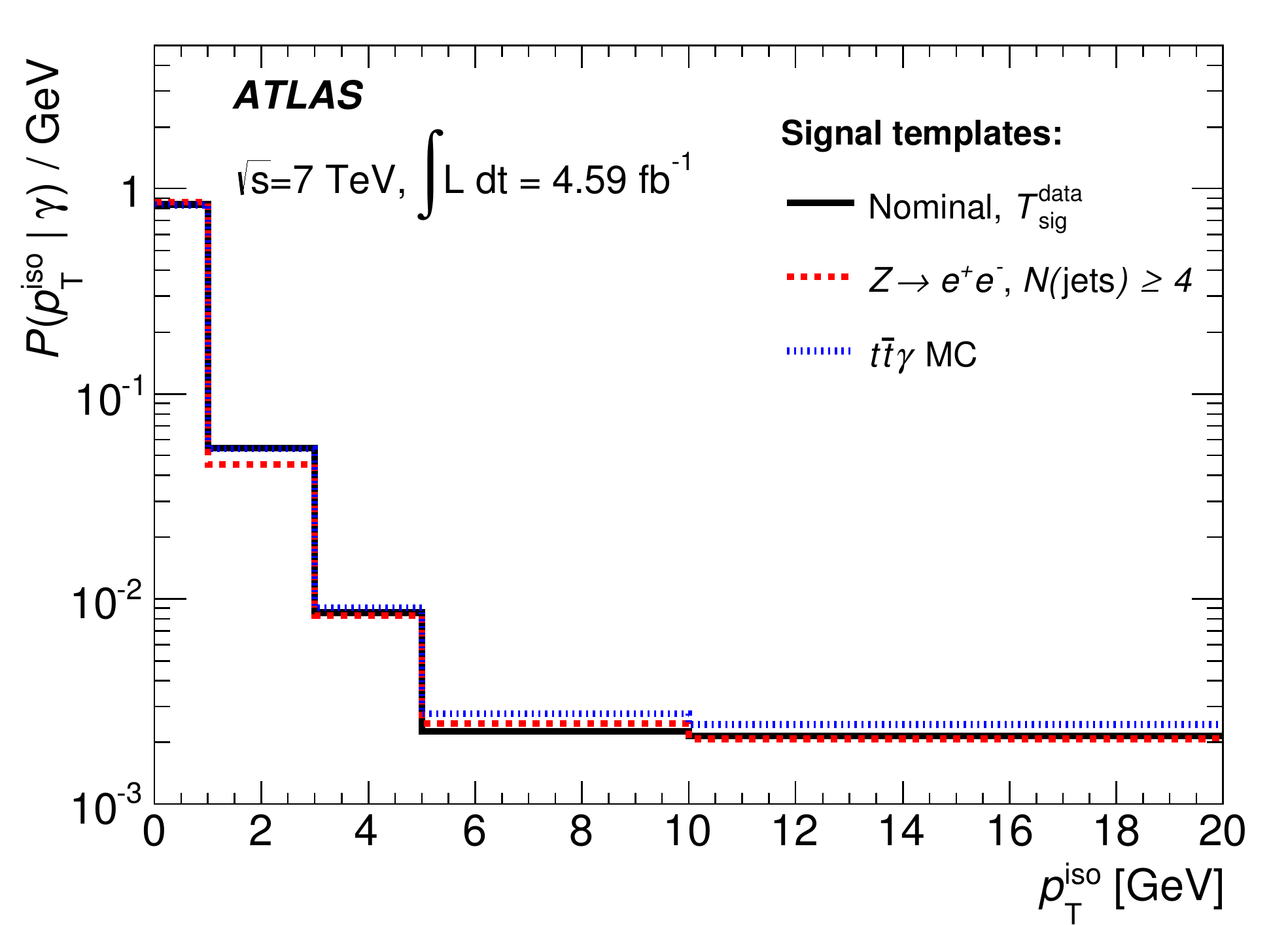}
\caption{Comparison of the nominal prompt-photon track-isolation ($\ptcone$)
  template with the template obtained from data using a \ZeejetsPar\ selection, and with the template obtained from 
  $\ttg$ simulation. 
The distributions show the probability $P(\ptcone |
  \gamma)$ of observing a photon in a given $\ptcone$ bin per GeV. The last bin contains any overflow.}
\label{fig:signal}
\end{figure}

Since electron and photon track-isolation distributions are expected
to be very similar, the electron template
$T_{\text{sig}}^{\text{data,}e}$ is extracted from the electron
\ptcone\ distribution in $\Zee$ candidate data events. The
prompt-photon template $T_{\text{sig}}^{\text{data}}$ is then derived
taking into account the differences between electron and photon
\ptcone\ distributions as well as differences between the $\Zee$ and
$\ttg$ event topologies, as photons from $\ttg$ events are less
isolated than electrons from $\Zee$ events. To obtain the
prompt-photon template, the electron \ptcone\ distribution in $\Zee$
candidate data events is corrected using weights ($w_i$) and templates
obtained from $\Zee$ ($T_{\text{sig},i}^{\text{MC,}e}$) and $\ttg$
($T_{\text{sig},i}^{\text{MC,}\gamma}$) Monte Carlo simulations in
twelve $\pt\times\eta$ bins (indexed by $i$):

\begin{small}
\begin{equation*}
T_{\text{sig}}^{\text{data}} = T_{\text{sig}}^{\text{data,}e} +
\sum_{i=\pt, \eta {\text{ bins}}}  w_i\left(T_{\text{sig},i}^{\text{MC,}\gamma}-T_{\text{sig},i}^{\text{MC,}e}\right).
\label{eq:signal_template}
\end{equation*}
\end{small}
The three $\pt$ bins are defined as $20~\GeV\le\pt<30~\GeV$,
$30~\GeV\le\pt<50~\GeV$, $\pt\ge50~\GeV$. The four $\eta$
bins are defined as $0.0 \le |\eta| < 0.6$, $0.6 \le |\eta| < 1.37$,
$1.52 \le |\eta| < 1.81$ and $1.81 \le |\eta| < 2.37$. The relative
weight for each bin $i$ is calculated from the photon $\et$ and $\eta$
spectra of the $\ttg$ Monte Carlo sample.
The prompt-photon
template, labeled as `Nominal', is shown in Fig.~\ref{fig:signal}. 
It is shown along with an electron \ptcone\ template obtained from
\ZeejetsPar\ candidate data events, and a prompt-photon \ptcone\ template
obtained directly from $\ttg$ Monte Carlo simulation. The latter two
templates are used to estimate systematic uncertainties on the
measured cross section due to the choice of the prompt-photon template.

\subsection{Background template}
\label{s:background_template}

Contributions from background sources with non-prompt photons are
described by a single template. This background template is extracted
from a multijet data sample by inverting requirements on photon shower
shape variables as described in
Sec.~\ref{s:background_derivation}. This set of events is referred
to as the `hadron-fake control region'. A correction is applied to
account for the prompt-photon contribution in the background template as
described in Sec.~\ref{s:contamination}.

\subsubsection{Derivation}
\label{s:background_derivation}

The hadron-fake control region is obtained from multijet events that
are required to have either at least two jets with \mbox{$\pt > 40~\gev$} and
at least two additional jets with \mbox{$\pt > 20~\gev$}, or at least five
jets with \mbox{$\pt > 20~\gev$}. Non-prompt photon candidates are identified
by inverting requirements on the electromagnetic shower
shapes~\cite{Aad:2010sp}. The background template shapes are
determined separately in the four photon $\eta$ bins and three photon
$\et$ bins defined in Sec.~\ref{s:signal_template}. The photon $\et$
distributions are consistent across different $\eta$ regions, so
$\eta$ and $\et$ dependencies of the background template are treated
separately.

To match the expected $\pt$ and $\eta$ distributions of non-prompt
photons in the signal region, these seven templates are weighted using
$\eta$ and $\pt$ distributions of non-prompt photon candidates in
$\ttbar$ candidate events in data. 
The resulting background template (labeled
as `Nominal template $T_{\text{bkg}}^{\text{data}}$') is shown in
Fig.~\ref{fig:DDMCComparison}.

\subsubsection{Prompt-photon contribution to the background template}
\label{s:contamination}

While the nominal background template is extracted using a data-based
procedure as described above, the
prompt-photon contamination in the background template is obtained using
a combination of data and Monte Carlo information.

Multijet simulation is used to obtain a Monte Carlo template modeling
the isolation distribution of hadrons misidentified as photons,
$T_{j\gamma}^{\text{MC}}$, by applying the same object and event
selection as for the nominal background template, as described in
Sec.~\ref{s:background_derivation}. A subset of the events used to
construct $T_{j\gamma}^{\text{MC}}$ is selected by the requirement
that those events do not contain any simulated true high-$\pt$
prompt photons. This subset is used to build a template
($T_{jj}^{\text{MC}}$) which models the isolation distribution of
hadrons misidentified as photons without any true prompt-photon
contribution.

\begin{figure}[!tb]
\centering
\includegraphics[width=0.5\textwidth]{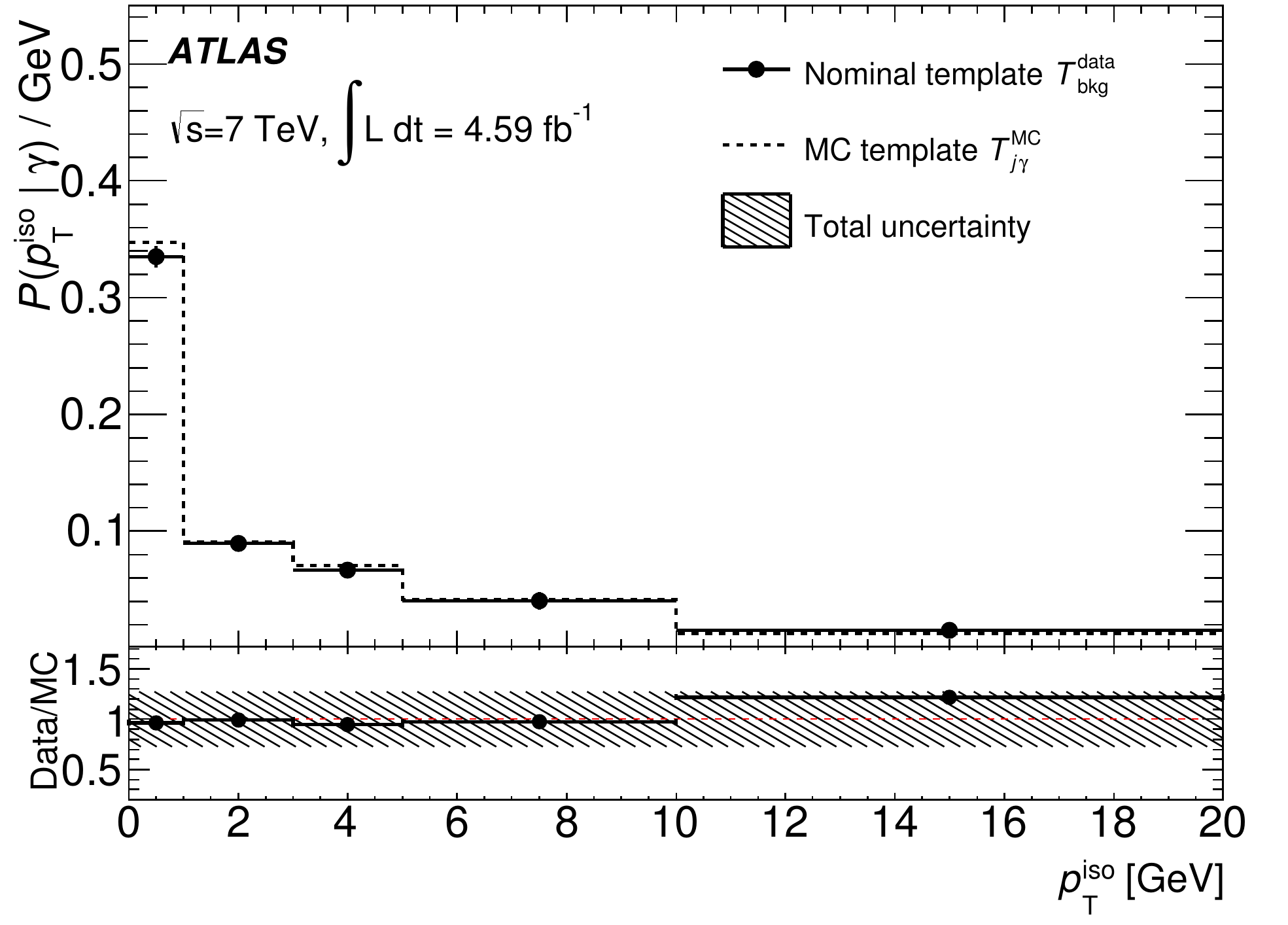}  
\caption{A comparison of  data-based $T_{\text{bkg}}^\text{data}$ and simulation-based $T_{j\gamma}^{\text{MC}}$ track-isolation background templates
  is shown in the upper panel. The distributions show the probability $P(\ptcone | \gamma)$ of observing
  a photon in a given \ptcone\ bin per GeV. The ratio of the two templates is
  shown in the lower panel. The hatched band shows the total
    uncertainty. The last bin contains any overflow.}
\label{fig:DDMCComparison}	
\end{figure}

Figure~\ref{fig:DDMCComparison} shows the comparison of
$T_{j\gamma}^{\text{MC}}$ to the data-based background template. 
The systematic uncertainty in each $\ptcone$ bin of
$T_{\text{bkg}}^\text{data}$ is assigned so that data
($T_{\text{bkg}}^\text{data}$) and simulation
($T_{j\gamma}^{\text{MC}}$) are in agreement. 
This uncertainty is conservatively
taken to be the same for all $\ptcone$ bins and is evaluated to
be 27\% on values of $T_{\text{bkg}}^\text{data}(\ptcone)$.

\begin{figure}[h!]
\centering	
\includegraphics[width=0.5\textwidth]{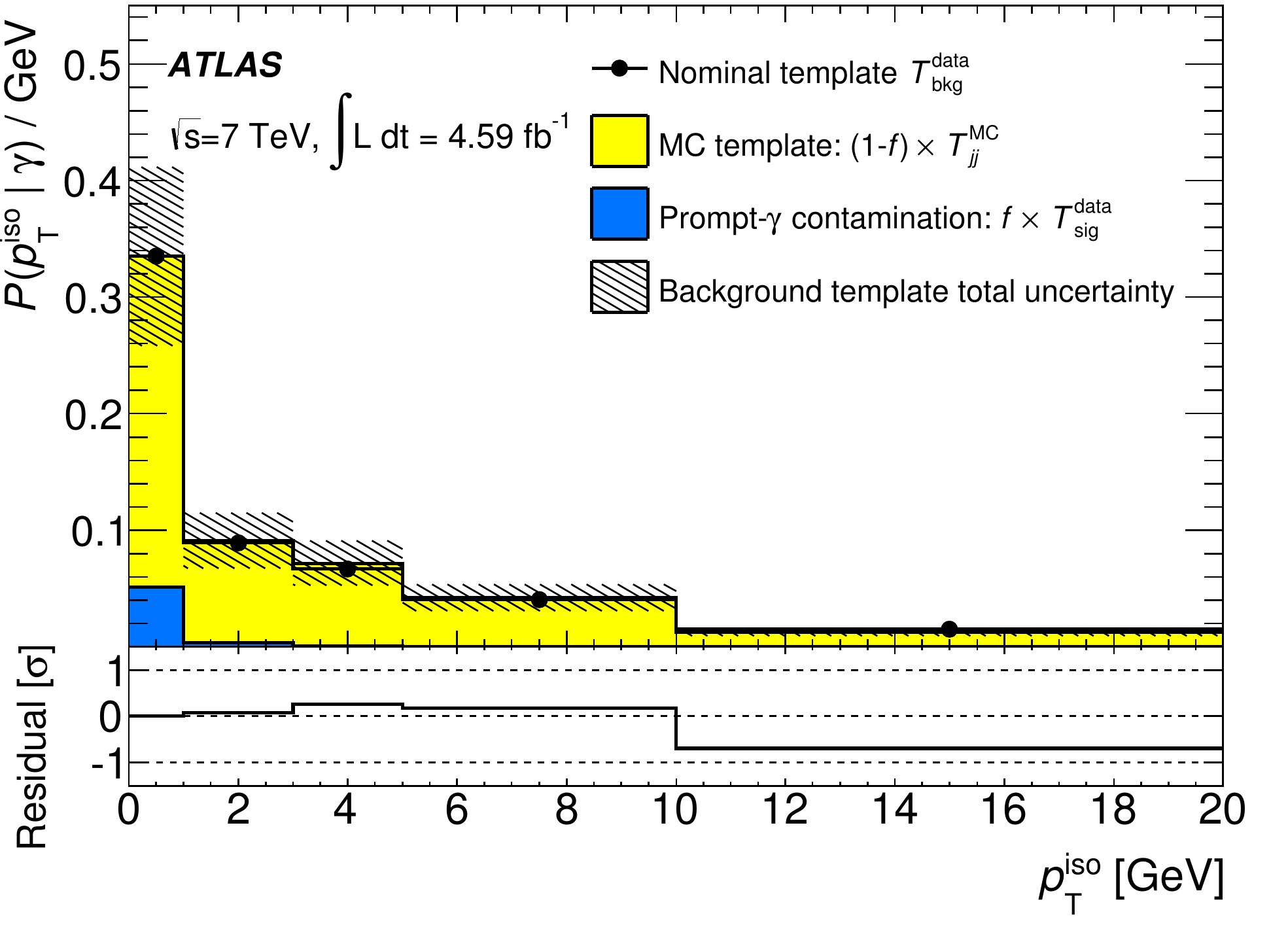}  
\caption{
Track-isolation background template distribution after maximization of the likelihood $L_f$
  defined in Eq.~\ref{eq:likelihood_fakescont} (top) and normalized residuals
  (bottom). The markers correspond to the nominal hadron background template. 
The stacked filled histograms represent the fraction of prompt photons
in the hadron-fake control region (obtained as $f\times
T_\text{sig}^{\text{data}}$) and the fraction of hadron-fakes 
(obtained from the simulation-based template as $(1-f)\times
T_{jj}^{\text{MC}}$) as given by the fit. 
The normalized residuals, shown in the
  bottom plot, are defined as the difference between the `Nominal
  template' and the sum of $ (1-f)\times T_{jj}^{\text{MC}}$ and
  $f\times T_\text{sig}^{\text{data}}$, divided by the total uncertainty $\sigma_\theta$. The last bin contains any overflow.}
\label{fig:FakesFitContamination}	
\end{figure}

The prompt-photon contamination is then extracted from data by
maximizing  the following extended likelihood function $L_f$,
representing the probability to observe $N$ data events in the hadron-fake 
control region given an expectation of $n_\text{exp}$:

\begin{multline}
L_f = \frac{n_{\text{exp}}^{N}}{N!}  e^{n_\text{exp}} \times \hat{\theta} \left [ \left(1-f\right)T_{jj}^\text{MC} + fT_\text{sig}^{\text{data}}\right]  \times \\ 
\mathcal{N}(\theta|\hat{\theta},\sigma_\theta),
\label{eq:likelihood_fakescont}
\end{multline}

\noindent where $T_\text{sig}^{\text{data}}$ is the prompt-photon
template and $f$ is the fraction of prompt photons. The parameter $\hat{\theta}$ is the nuisance parameter modeling the systematic uncertainty due to the differences between $T_{\text{bkg}}^\text{data}$ and $T_{j\gamma}^{\text{MC}}$.
The fraction of prompt photons is distributed according to a
Gaussian pdf $\mathcal{N}(\theta|\hat{\theta},\sigma_\theta)$ with
mean $\hat{\theta}=1$ and width $\sigma_\theta=27\%$. The result of
the fit is shown in Fig.~\ref{fig:FakesFitContamination}, and $f$ is
determined to be $( 6.1 ^{+1.7}_{-0.9}) \times 10^{-2}$. 
The uncertainties are obtained at the 68\% confidence level (CL) by
constructing the confidence belt with the Feldman--Cousins
technique~\cite{feldman-cousins} using pseudoexperiments.

Finally, the signal contamination in the background template is
included in the general likelihood by means of a nuisance parameter
$\afake$ modeling the strength of the correction:

\begin{small}
\begin{equation}
 T_{\text{bkg}}^\text{corr}= \left(\frac{1}{1-\alpha_\text{fake}\cdot f}\right) \left[ T_{\text{bkg}}^\text{data} - \alpha_\text{fake}\cdot f \times T_\text{sig}^{\text{data}} \right]. \nonumber
\end{equation}
\end{small}
\noindent
The strength factor $\afake$ is constrained to $1$ by a Gaussian pdf with
width $\sigma_\alpha = 28\%$ corresponding to the largest of the estimated asymmetric uncertainties on $f$. 
It is then determined from the general likelihood fit in a data-based way.

\section{Prompt-photon backgrounds}
\label{s:backgrounds}

To identify prompt-photon and isolated-electron background contributions to the
events selected in the $\ttg$ analysis, data-based methods and Monte
Carlo simulation are used. These background estimates are summarized in
Table~\ref{tab:backgrounds_summary} and described below.

\begin{table}[!bhtp]
\caption{Estimates of the number of selected events with prompt
  photons, or electrons misidentified as photons, from various
  backgrounds to $\ttg$ production, including
  statistical and systematic uncertainties.}  
\centering
\begin{tabular}{c|c|c}
\hline 
\hline 
Background source  & Electron channel & Muon channel   \\
\hline
$e\rightarrow\gamma$ misidentification & 29.4  $\pm$ 3.0~~ & 41.5  $\pm$  4.6~~ \\
Multijet + $\gamma$                    &  1.4 $\pm$ 1.2  &  1.9 $\pm$ 1.1 \\
$W\gamma$ + jets                       &  5.4 $\pm$  1.9  & 15.6  $\pm$ 4.4~~ \\
Single-top-quark + $\gamma$          &  1.8 $\pm$ 0.3  &  3.8 $\pm$ 0.4 \\
$Z\gamma$ + jets                  &  2.3  $\pm$ 1.6  &  4.2  $\pm$  3.1 \\
Diboson                                &  0.1 $\pm$ 0.1  &  0.4 $\pm$ 0.1 \\ 
\hline
\hline
\end{tabular}
\label{tab:backgrounds_summary}
\end{table}

\subsection{Electron misidentified as a photon}
\label{s:ephotonfakes}

The contribution from events with an electron misidentified as a
photon is estimated using data by applying the $e\rightarrow\gamma$
misidentification rate to $t\bar{t}+e$ candidate events. The
measurement of this misidentification probability and cross-checks of
the method are described below.

The sample of events with an electron and a photon approximately
back-to-back in the transverse plane (in $\phi$) with an
electron--photon invariant mass $\meg$ close to the $Z$-boson mass is
dominated by $\Zee$ decays in which one of the electrons radiates a
high-$\et$ photon while traversing detector material. The probability
for an electron to be misidentified as a photon is determined in data
as a function of the electron transverse momentum and pseudorapidity
using the $e\gamma$ and $\ee$ mass distributions.  One electron (tag)
is required to match the single-electron trigger. Another
electromagnetic object (probe), an electron or photon, is then required
to be present and give a di-object mass with the tag close to the
$Z$-boson mass.  The $e\gamma$ and $\ee$ mass distributions are fit
with the sum of a Crystal Ball~\cite{Oreglia:1980cs,Gaiser:1982yw}
function (for the signal part) and a Gaussian function (for the
background part) to obtain the numbers of $ee$ and $e\gamma$ pairs,
$N_{ee}$ and $N_{e\gamma}$, to which several pairs per event can
enter. The probability of an electron being misidentified as a photon
is measured in $\eta$ and $\pt$ bins as $f_{e\rightarrow\gamma} =
N_{e\gamma} / N_{ee}$.

The nominal selection for the signal $\ttg$ region is modified by
replacing the photon requirement by an extra-electron
requirement.  This extra electron ($e_{\rm f}$) must fulfill the photon kinematic
selection, $\et(e_{\rm f})>20~\GeV$ and $|\eta (e_{\rm f})|<2.37$, excluding the
transition region between the barrel and endcap calorimeters at
1.37$<|\eta(e_{\rm f})|<$1.52. To estimate the contribution from an electron
misidentified as a photon, these `$\ttbar + e$' events are reweighted
according to the probability of the extra electron being misidentified as a
photon. 
This procedure gives $29.4 \pm 3.0$ and $41.5 \pm 4.6$ events in the
electron and muon channels respectively.

The misidentification probability $f_{e\rightarrow\gamma}^{\rm MC}$ is
also estimated in $\Zee$ Monte Carlo simulation, so that a closure
test can be performed.  The number of background events in
simulation that pass the $\ttg$ event selection is estimated using
generator-level information about how the photon is produced. These
events are weighted with the data-to-simulation correction factors
$s_{e\gamma} = f_{e\rightarrow\gamma} / f_{e\rightarrow\gamma}^{\rm
  MC}$ found typically to be within 10\% of unity. This estimate is
found to be in agreement with reweighting the events that pass the
`$\ttbar + e$' event selection in Monte Carlo simulation according to
$f_{e\rightarrow\gamma}$, i.e. effectively using the data-based
approach in the Monte Carlo simulation.

\subsection{Multijet + photon}
\label{s:qcd}

The background contribution from multijet events with associated
prompt-photon production is estimated using the data-based matrix
method discussed in more detail in Ref.~\cite{Aad:2010ey}. In this
method, two sets of lepton selection criteria are defined. The `tight'
selection criteria are used to identify leptons in $\ttg$ candidate
events. In the `loose' selection criteria the lepton isolation
requirements are disregarded, and looser identification 
requirements~\cite{Aad:2010sp} are applied for electrons. 

The number of selected $\ttg$ candidate events is expressed as a sum
of those with prompt leptons and those with `fake leptons' (non-prompt
leptons or hadrons misidentified as leptons). Identification
efficiencies for prompt leptons are measured in $\Zll$ ($\ell \equiv
e$, $\mu$) data candidate events, whereas the efficiency for fake
leptons to be identified as `tight' leptons is measured in a multijet
data sample. The number of $\ttg$ candidate events with at least one
non-prompt lepton candidate is estimated using this
information~\cite{Aad:2010ey}.

A template fit to the photon \ptcone\ distribution is used to
determine the prompt-photon fraction in selected `multijet + $\gamma$'
events. The `multijet + $\gamma$' event selection is similar to the
$\ttg$ selection except that `loose' lepton identification criteria
are used instead of the `tight' criteria. Assuming that the
prompt-photon fraction does not depend on the lepton identification
criteria (`loose' or `tight'), this prompt-photon fraction is then
used to estimate the contribution of the multijet + prompt-photon
process to the $\ttg$ event selection. 
This results in 1.4 $\pm$ 1.2 and 1.9 $\pm$ 1.1 events expected for
the electron and muon channels respectively.

\subsection{$W\gamma$ + jets production}

Background from $\Wgjets$ production is estimated by extrapolating the
number of $\Wgjets$ candidate events in data from a control region
(CR) to the $\ttg$ signal region (SR) using $\Wgjets$ Monte Carlo
simulation~\cite{Berends:1990ax}. In the control region the lepton,
photon, $\met$ and $\mtw$ selection criteria are the same as in the
nominal $\ttg$ selection. To enrich the control region in $\Wgjets$,
events are required to have one, two or three jets, and a $b$-tagging
veto is applied.

To estimate the prompt-photon contribution, it is assumed that the
fraction of prompt photons is the same in the CR and SR. 
To verify this assumption, a template fit to the photon
\ptcone\ distribution is performed, and the prompt-photon fraction in
data and simulation is found to be independent of the jet
multiplicity.

To suppress the $\Zjets$ background contribution in the CR, the $\meg$
requirement is extended to \mbox{$|\meg-m_Z|>15~\GeV$}. The multijet +
$\gamma$ contribution to the $\Wgjets$ background in the CR is
estimated using the matrix method as described in
Sec.~\ref{s:qcd}. The number of $\Wgjets$ events with prompt
photons in the CR is estimated using a template fit to the photon
\ptcone\ distribution.

Other contributions to the $\Wgjets$ CR are estimated using simulation, where events are separated into two classes, one
with a prompt photon, the other with an electron misidentified as a
photon. To obtain the $e\rightarrow\gamma$ contribution, the
$s_{e\gamma}$ correction factors (Sec.~\ref{s:ephotonfakes}) are
used. A comparison of data and expectation in the CR is presented in
Table~\ref{t:wgamma}.

\begin{table}[!tbhp] 
\caption{Data and simulated background yields in the
$\Wgjets$ data control region. The number of events with a prompt
photon in data (labeled as `Events with prompt $\gamma$') is estimated
from the total number of $\Wgjets$ candidate events in the control
region (labeled as `$\Wgjets$ control region') using template
fits. Background yields are estimated using Monte Carlo (MC)
simulation, except for the multijet + $\gamma$ yield.  The resulting
number of $W\gamma$ candidate data events, as well as the MC
prediction for the number of $W\gamma$ events are shown. To obtain the
$\Wgjets$ background to the $\ttg$ selection, the number of $W\gamma$
candidate data events is extrapolated into the signal region using
Monte Carlo simulation. The uncertainties include both the statistical
and systematic uncertainties.}
\centering
\begin{tabular}{c|c|c}
\hline
\hline
 & {Electron channel} & {Muon channel} \\
\hline
$\Wgjets$ control region & 3410 &  8394 \\ 
\hline
Events with prompt $\gamma$ & 2412 & 5540 \\
\hline
\ttg & 82  $\pm$  16  &161 $\pm$  32  \\
$\Zjets$ & 160 $\pm$ 90  &  620 $\pm$ 330  \\ 
Diboson & 13  $\pm$ 3 &  26  $\pm$ 7 \\ 
Single-top-quark & 9  $\pm$ 2  &  20  $\pm$ 5   \\
$e\rightarrow\gamma$ misidentification &380  $\pm$ 110 &  330  $\pm$ 40  \\ 
Multijet + $\gamma$ & 60   $\pm$ 30 &  350  $\pm$ 70 \\ 
\hline
Total background & 700  $\pm$ 140 & 1510 $\pm$ 340 \\ 
\hline\hline
$W\gamma$ estimate &1710 $\pm$ 180 & 4030 $\pm$ 390 \\
\hline
$W\gamma$ MC expectation &1860  $\pm$ 200 & 3930  $\pm$ 390\\
\hline\hline
\end{tabular}
\label{t:wgamma}
\end{table}

The number of $\Wgjets$ candidate events in the CR ($\le 3$ jets) is
extrapolated to the jet multiplicity of the SR, $\ge 4$
jets~\cite{Aad:2010ey}. 
To extrapolate from the $\Wgjets$ event selection, which has a $b$-tagging veto, to the SR,
the heavy-flavor quark content is studied in data in events with a $W$ boson and two
jets. The heavy-flavor quark content is then extrapolated from the
$W\gamma$ + 2-jets region into the SR using the $\Wgjets$ simulation~\cite{Berends:1990ax,Aad:2010ey}. This extrapolation
accounts for the difference in flavor composition between the $\Wg + 2$-jet
and $\Wg + \ge 4$-jet samples as well as for differences in the
per-flavor event tagging probabilities, which may lead to different
event rates after $b$-tagging. The $\Wgjets$ background estimate is
5.4 $\pm$ 1.9 and 15.6 $\pm$ 4.4 events for the electron and muon
channels respectively.

Monte Carlo modeling uncertainties in the estimate of the background from 
$W\gamma$ + jets production include
contributions from the estimated number of events with electrons misidentified as 
photons (which is known to 10\%) and from cross section uncertainties 
(e.g. a 48\% uncertainty for $Z$+jets contributions, which corresponds to the error on the normalization of $Z$+jets 
in the four-jet bin from the Berends--Giele scaling~\cite{Berends:1990ax}).

\subsection{Other background sources}

The single-top-quark, $Z$+jets, and diboson contributions are
estimated from simulation and normalized to theoretical
calculations of the inclusive cross sections.

The single-top-quark production cross section is normalized to the NLO+NNLL
prediction: the $t$-channel to $64.6{}^{+2.6}_{-1.7}\,{\rm
  pb}$~\cite{Kidonakis:2011wy}, the $s$-channel to $4.6{}\pm 0.2\,{\rm
  pb}$~\cite{Kidonakis:2010tc}, and the $Wt$-channel to $15.7{} \pm
1.2 \,{\rm pb}$~\cite{Kidonakis:2010ux}. The $Z$+jets background is
normalized to the NNLO QCD calculation for inclusive
$Z$~production~\cite{Anastasiou:2003ds} and the diboson
background is normalized to the NLO QCD cross section
prediction~\cite{Campbell:2011bn}.

\section{Systematic uncertainties}
\label{s:systematics}

Systematic uncertainties may affect the shapes of the \ptcone\ prompt-photon
and background templates, the estimates of background components with prompt
photons and with electrons misidentified as photons, as well as the
efficiencies, acceptance factors and the luminosity.

The total effect of each systematic uncertainty on the cross section is
evaluated using ensemble tests. For each systematic uncertainty $i$,
pseudodata are generated from the full likelihood while keeping all
parameters fixed to their nominal values except for the nuisance
parameter corresponding to the systematic uncertainty
source. For each set of pseudodata, a template fit is performed allowing all
parameters of the likelihood (nuisance parameters, signal cross
section) to vary.
The distribution of cross sections obtained form a Gaussian pdf with a
width that gives
 the uncertainty in the cross section due to the $i$-th systematic uncertainty.
This method provides an estimate of the effect of each
uncertainty on the cross section as shown in
Table~\ref{tab:systematics}. Uncertainties obtained with this method
are by construction symmetric. 
All systematic uncertainties are described in the following.

\begin{table}[thbp]
\caption{Summary of systematic uncertainties on the $\ttg$ fiducial cross section, $\sttgfid$.}
\centering
\begin{tabular}{c | c }
\hline 
\hline 
Uncertainty source & Uncertainty [\%] \\
\hline 
Background template shapes  & 3.7 \\ 
Signal template shapes         & 6.6 \\ 
\hline 
Signal modeling                & 8.4 \\ 
\hline                         
Photon modeling                & 8.8 \\
Lepton modeling                & 2.5 \\ 
Jet modeling                   & 16.6 \\
$b$-tagging                    & 8.2   \\ 
$\met$ modeling                & 0.9 \\
Luminosity & 1.8 \\ 
\hline                         
Background contributions       & 7.7 \\
\hline 
\hline 
\end{tabular}
\label{tab:systematics}
\end{table}

\subsection{Template shapes}
\label{s:sys_shapes}

The contribution to the systematic uncertainty on $\sttgfid$ due to
the template shape modeling amounts to 7.6\% in total. 
Of this, the background template shape modeling uncertainty amounts to
3.7\% of the cross section, and the prompt-photon template uncertainty
amounts to 6.6\%.

The prompt-photon template shape systematic uncertainty is estimated with
pseudoexperiments by replacing the nominal prompt-photon template with 
alternative templates shown in Fig.~\ref{fig:signal}: 
(a) an electron \ptcone\ template obtained
from 
\ZeejetsPar\ candidate data events (4.1\%
systematic uncertainty is obtained) and (b) a prompt-photon \ptcone\ template
obtained directly from $\ttg$ Monte Carlo simulation (6.6\% systematic
uncertainty is obtained). 
The larger of the two
uncertainties is used as the systematic uncertainty.

The systematic uncertainty associated with the reweighting of the
background template is estimated by varying within their uncertainties the non-prompt photon
$\pt$- and $\eta$-distributions that are used for reweighting. The effect of this systematic uncertainty on the cross
section measurement is found to be negligible. To estimate the
systematic uncertainty due to the amount of prompt-photon
contamination in the background template (as described in
Sec.~\ref{s:background_template}), the corresponding nuisance
parameter $\afake$ is sampled using a Gaussian pdf with a width of
$\sigma_{\afake}=28$\% corresponding to its estimated uncertainty. The
systematic uncertainty on the cross section is estimated to be 3.7\%.
All template-shapes uncertainties are taken as fully correlated between the electron channel and the muon channel.

\subsection{Signal modeling}
\label{s:sys_signal}

The uncertainty on the $\ttg$ cross section (as defined in
Sec.~\ref{s:fiducial}) due to the modeling of the signal is estimated
to be 8.4\%. The estimate is obtained by varying the selection
efficiency with respect to the nominal $\ttg$ Monte Carlo sample which
includes event migrations into and out of the fiducial region. 
This uncertainty
includes a comparison of \madgraph\ with \whizard\ (1.7\%), as well as a
comparison of the \madgraph\ $\ttg$ samples 
with different QED FSR settings (3.4\%) as explained in Sec.~\ref{s:samples}.
The renormalization and factorization scales are also varied, leading to an uncertainty of 1.1\%.
To assess the effect of different parton shower models, predictions
from the \madgraph+\herwig\ sample are compared to predictions from
the \madgraph+\pythia\ sample, leading to an uncertainty of 7.3\%.
In addition, studies of $t\bar{t}$ samples with varied color reconnection (0.2\%) and underlying event (0.9\%) settings lead to small contributions. 
The uncertainty associated with the choice of the CTEQ6L1 PDF set is evaluated
from an envelope of calculations
using the PDF4LHC prescription~\cite{Botje:2011sn}
by reweighting the CTEQ6L1 LO PDF used in the generation of the $\ttg$
\whizard\ sample with
MSTW2008~\cite{Martin:2009iq,Martin:2009bu}, 
CT10~\cite{Lai:2010vv,Gao:2013xoa} 
and NNPDF2.0~\cite{Ball:2012cx} NLO PDF sets and amounts to 1.1\%.
All signal-modeling uncertainties are taken as fully correlated between the electron channel and the muon channel.

\subsection{Detector modeling}
\label{s:sys_detector}

The systematic uncertainty on the cross section due to 
photon modeling is 8.8\%. It is estimated from the
photon identification (7.3\%)~\cite{ATLAS:2012ar}, the electromagnetic
energy scale (2.7\%) and the resolution (4.0\%) systematic
uncertainties~\cite{ATLASElectronPerformance2010}.

The systematic uncertainty on the cross section due to 
lepton modeling is 2.5\%. It is
estimated separately for the electron and muon channels from the
lepton trigger (0.3\% and 1.7\%), reconstruction (0.5\% and 0.4\%) and
identification (1.2\% and 1.0\%) efficiency uncertainties, as well as
from those on the energy scale (0.3\% and 0.3\%) and resolution (0.1\%
and 0.7\%).

The systematic uncertainty on the cross section due to 
jet modeling is 16.6\%. It is estimated
taking into account the following contributions. The largest effect
comes from the energy scale (15.0\%) uncertainty which is estimated by
combining information from the single-hadron response measured with
in-situ techniques and with single-pion test-beam measurements~\cite{Aad:2014bia}.  
The jet energy resolution (6.5\%) uncertainty is estimated by
smearing the jets in simulation by 
the uncertainty as measured with the dijet balance and bisector
techniques~\cite{Aad:2012ag}.  
The uncertainty on jet reconstruction
efficiency (1.0\%), which is defined relative to jets built
from tracks reconstructed with the ID, is also considered~\cite{Aad:2011he}. 
The jet vertex fraction uncertainty is found to be 2.6\%.

The systematic uncertainty on the cross section due to 
$b$-tagging modeling is 8.2\%. It is
dominated by the contribution due to the efficiency
(8.1\%)~\cite{ATLAS-CONF-2012-043} with a small contribution due to
the mistag probability (1.1\%)~\cite{ATLAS-CONF-2012-040}.

Systematic uncertainties on the energy scale and resolution of
leptons, jets and photons are propagated to $\met$. Additional $\met$
uncertainties~\cite{Aad:2012re} also taken into account are contributions from low-$\pt$ jets and from
energy in calorimeter cells that are not included in the
reconstructed objects (0.3\%), as well as any dependence on pile-up (0.9\%).

All detector-modeling systematic uncertainties except for the lepton-modeling uncertainties are taken as fully correlated between the electron channel and the muon channel.
The lepton-modeling uncertainties are taken as uncorrelated between channels.

The effect of the luminosity uncertainty on the cross section amounts to 1.8\%~\cite{Aad:2013ucp}.

\begin{figure*}[!t]
\centering	
\includegraphics[width=0.5\textwidth]{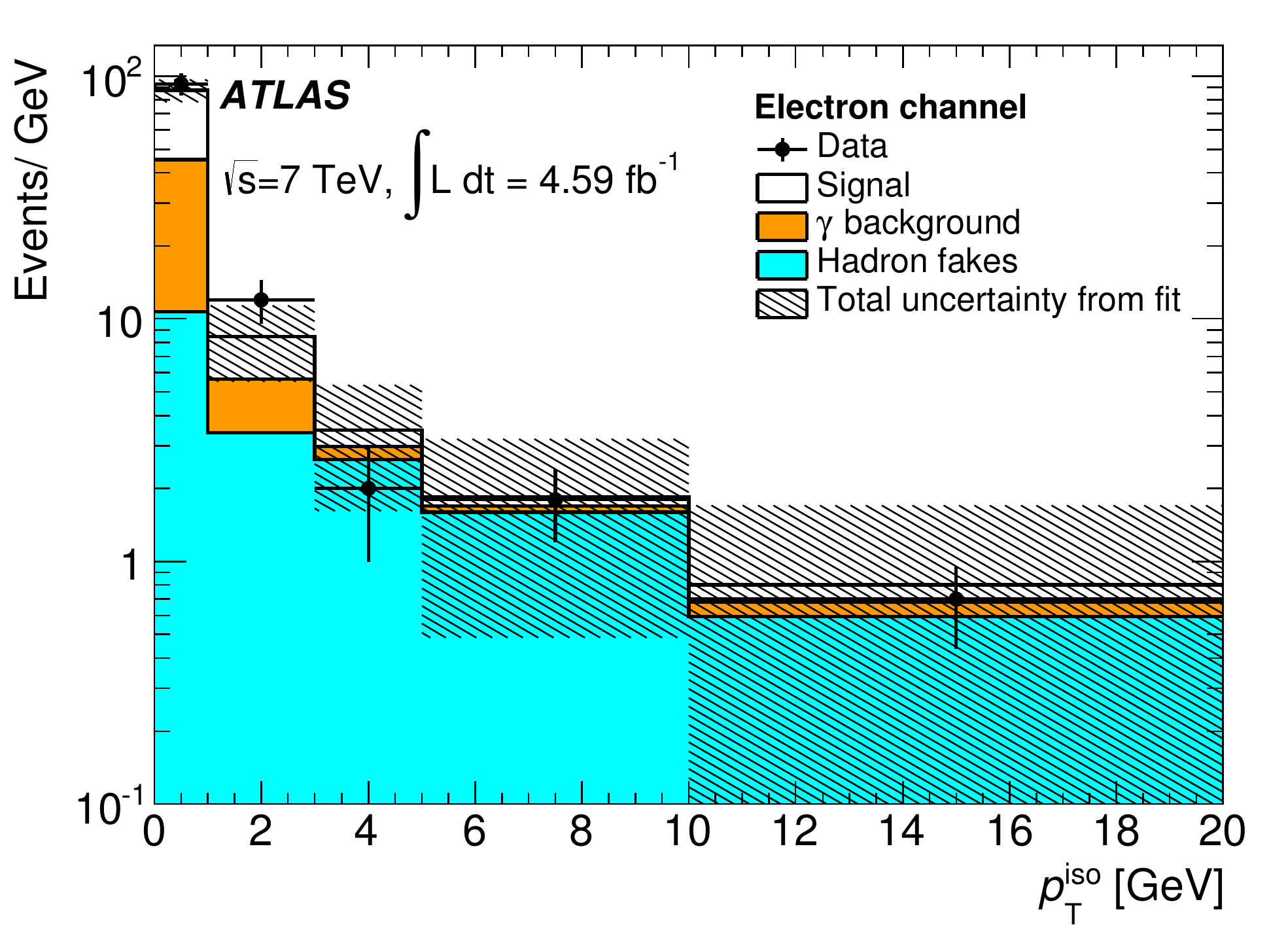}\hspace{-0.2cm}
\includegraphics[width=0.5\textwidth]{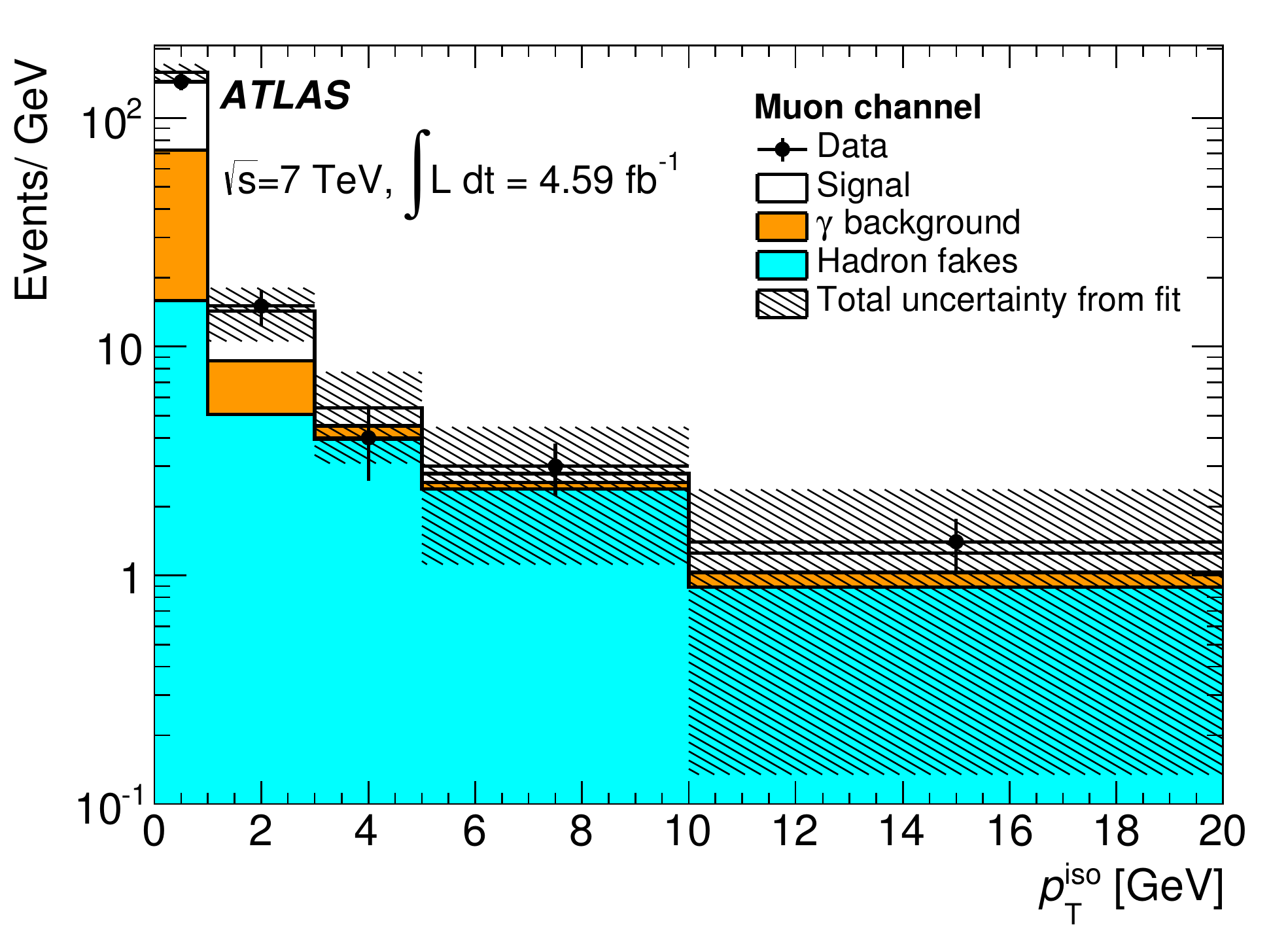}\vspace{-0.3cm}
\caption{Results of the combined likelihood fit using the track-isolation (\ptcone) distributions as the discriminating variable for the
  electron (left) and muon (right) channels. The contribution from $\ttg$ events is labeled as `Signal', prompt-photon background is labeled `$\gamma$ background', the contribution from hadrons misidentified as photons (as estimated by the template fit) is labeled as `Hadron fakes'.}
\label{fig:IsoFitEl}	
\end{figure*}

\subsection{Background contributions}
\label{s:sys_backgrounds}

The total systematic uncertainty originating from the non-$\ttg$ background
contributions with prompt photons or electrons misidentified as
photons is estimated to be 7.7\%. 
This uncertainty includes
 the following: electrons misidentified as photons (5.0\%), 
$W\gamma$+jets (5.4\%), as well as multijet + photon (1.5\%), 
$Z\gamma$+jets (1.3\%), diboson (0.4\%) and single-top-quark (0.4\%) processes. 
The various sources of uncertainty on the background estimates quoted above are described in the following paragraphs.

For background estimates obtained
using simulation, uncertainties on the cross section
predictions are taken into account. Cross section systematic uncertainties are considered as fully correlated between the electron 
and the muon channels. However, the corresponding statistical uncertainty is taken as uncorrelated.  For $Z\gamma$+jets, single-top-quark 
and diboson contributions the cross section systematic uncertainty is negligible with respect to the statistical uncertainty.

The systematic uncertainty on the probability of an electron to be
misidentified as a photon as described in Sec.~\ref{s:ephotonfakes}
is obtained by varying the fit functions and the $ee$ and $e\gamma$
mass windows in $\Zee$ candidate events in data. This uncertainty is
estimated to be about 10\% of the background estimate and it is taken as fully correlated between the electron channel and the muon channel.

For the multijet + photon background described in Sec.~\ref{s:qcd},
the uncertainty is about 90\% for the electron channel and 60\% for
the muon channel. It is dominated by the statistical uncertainty due to the small
number of events in the data samples and the systematic uncertainties on the matrix method (50\%
for the electron channel and 20\% for the muon channel)~\cite{Aad:2010ey}.  Those uncertainties are taken as uncorrelated between the two channels.

The systematic uncertainties on the $W\gamma$+jets background are
dominated by the extrapolation from the control region (dominated by
$W\gamma$+jets) to the signal region due to different event topologies
in the two regions in terms of the total number of jets and the number
of heavy-flavor jets. The uncertainties due to the extrapolation are
27\% in the electron channel and 23\% in the muon channel and are
dominated by the uncertainty on the knowledge of the flavor
compositions of the $W$+jets events and the overall $W$+jets
normalization for different jet
multiplicities~\cite{Berends:1990ax,Aad:2010ey}. Those uncertainties are taken as fully correlated between the electron channel and the muon channel. 
The statistical uncertainty on the number of events in the $W\gamma$+jets control region is taken as uncorrelated between the two channels.
Systematic uncertainties on the multijet+photon contribution to the
$W\gamma$+jets event selection, as well as uncertainties on Monte
Carlo modeling of $\ttbar$, $Z$+jets, diboson, and single-top-quark
processes are taken into account~\cite{Aad:2012qf}. 

\section{Results}
\label{s:results}

Totals of \ttgel\ and \ttgmu\ $\ttg$ candidate data events are
observed in the electron and muon channels respectively. The numbers
of background events extracted from the combined likelihood fit are
\mbox{$\bkgttgel$} for the electron channel and \mbox{$\bkgttgmu$} for
the muon channel. The numbers of $\ttg$ signal events are determined
to be $\sigCombel \pm \sigCombelErr$ and \mbox{$\sigCombmu \pm
  \sigCombmuErr$}. The results include statistical and systematic
uncertainties. These numbers are summarized in
Table~\ref{tab:fitResults}, and the \ptcone\ distributions are shown
in Fig.~\ref{fig:IsoFitEl}.

\begin{table}[!thbp]
\caption{Number of $\ttg$ signal and background events
  extracted from the likelihood fit, which is performed for the
  electron and muon channels simultaneously. The uncertainties are
  statistical and systematic. The total number of $\ttg$ candidate
  events observed in data is also shown.}
\begin{tabular}{ l |   c |  c | c}
\hline
\hline
Contribution & Electron chan.  & Muon chan. &  Total \\
\hline
Signal & \sigCombel\  $\pm$ \sigCombelErr & \sigCombmu\ $\pm$ \sigCombmuErr & \sigComTot~$\pm$~\sigComErr\\
\hline
Hadrons           & \hadFakesCombel\  $\pm$ \hadFakesCombelErr & \hadFakesCombmu\ $\pm$ \hadFakesCombmuErr & \hadFakesComb ~$\pm$~\hadFakesCombErr \\ 
Prompt photons & \promptPhotonsCombEl\  $\pm$ \promptPhotonsCombElErr & \promptPhotonsCombMu\ $\pm$ \promptPhotonsCombMuErr & \promptPhotonsComb~$\pm$~\promptPhotonsCombErr \\
\hline
Total background  & $\bkgttgel$ & $\bkgttgmu$ & \bckCombTot~$\pm$~\bckCombTotErr \\
\hline 
Total & 131 $\pm$  30 & 220 $\pm$ 48  & 351 $\pm$ 59 \\
\hline
Data candidates  & \ttgel & \ttgmu & \multicolumn{1}{c}{362} \\
\hline
\hline
\end{tabular}
\label{tab:fitResults}
\end{table}

Using the asymptotic
properties~\cite{Wilks:1938dza} of the likelihood model, the test
statistic for the no-signal hypothesis is extrapolated to the likelihood
ratio value observed in data ($\LLratioObsVal$) to determine the p-value of
$\pvalue$. The process $\ttg$ in the lepton-plus-jets
final state is observed with a significance of
$\mysignificance\sigma$ away from the no-signal hypothesis.

\begin{figure}[!bhtp]
\centering	
\includegraphics[width=0.5\textwidth]{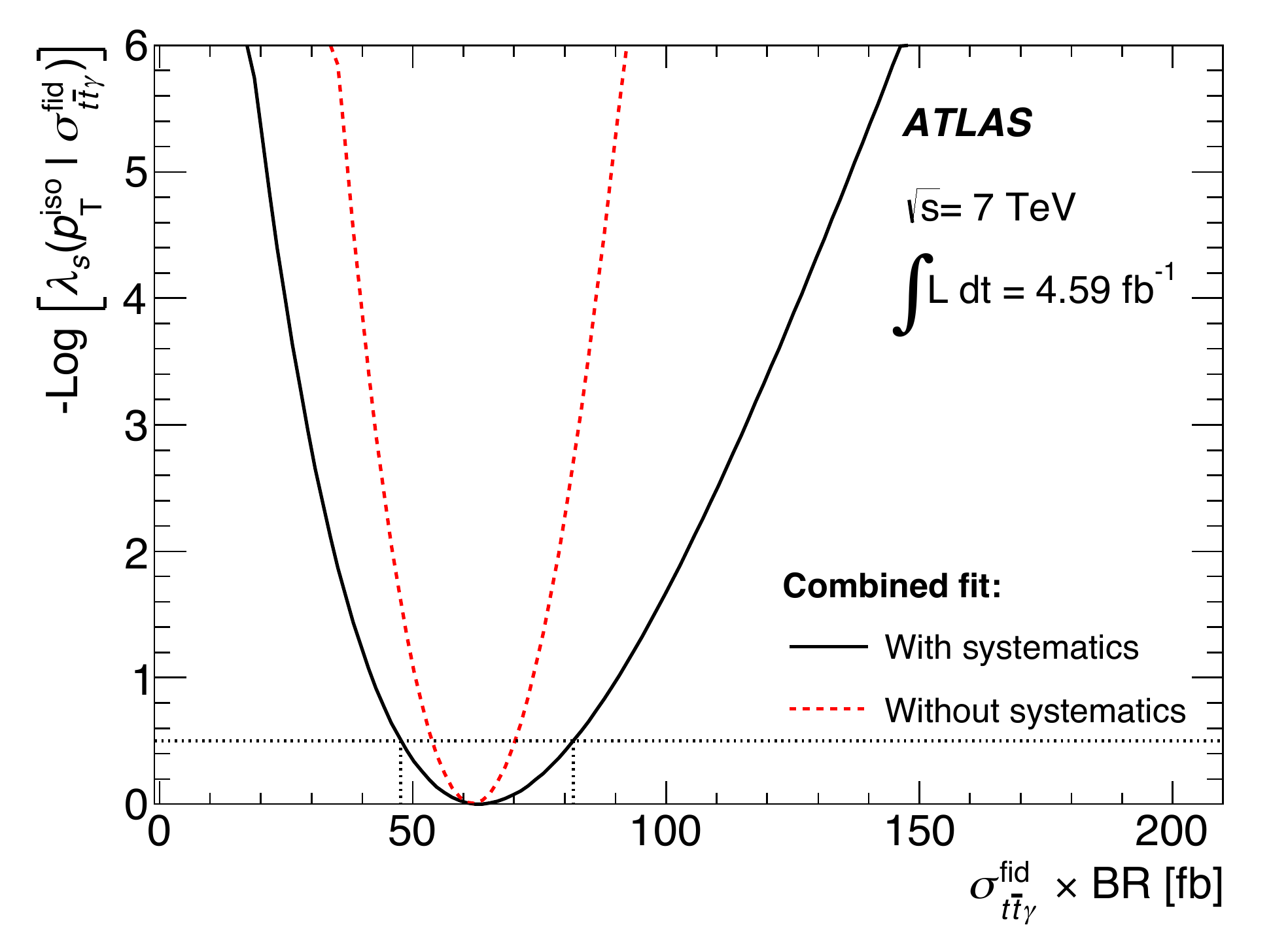}
\caption{Negative logarithm of the profile likelihood as a function of the \ttg\ fiducial cross section $\sttgfid\times \text{BR}$ with (solid line) and without (dashed line) free nuisance parameters associated with the systematic uncertainties. 
\thisdottedlinerules 
}
\label{fig:LikelihoodNoNuis}	
\centering
\end{figure}

The $\ttg$ fiducial cross section together with its total uncertainty
is obtained from the profile likelihood ratio fit to be $\xsectot$. 
The total systematic
uncertainty is extracted from $\sqrt{(\sigma_{\rm{syst\oplus stat}})^2 - \sigma^2_{\rm{stat}}-\sigma^2_{\rm{\mathcal{L}}}} =
\Systerrargh$~fb, where $\sigma_{\rm{\mathcal{L}}}$ is the luminosity uncertainty; $\sigma_{\rm{stat}}$ is the pure
statistical uncertainty, evaluated from the profile likelihood without
including nuisance parameters; $\sigma_{\rm{syst \oplus stat}}$ is the total
uncertainty extracted from the 68\% CL of the profile likelihood fit
(including nuisance parameters), as shown in
Fig.~\ref{fig:LikelihoodNoNuis}.

The $\ttg$ fiducial cross section times BR per lepton flavor, as
defined in Sec.~\ref{s:fiducial}, is determined to be $\xsecexpfb$, 
where BR is the $\ttg$ branching ratio in the single-electron or single-muon final state. 
Good agreement is found with the predicted cross
sections~\cite{Melnikov,Melnikov:private} of $\xsectheorfb$ and
$\xsectheorfbMG$ obtained from the \whizard\ and \madgraph\ Monte
Carlo generators respectively and then normalized by the corresponding
NLO/LO $K$-factors. In addition, the cross section measurements are
performed separately in the electron and muon channels and give
$\xsecexpelfb$ and $\xsecexpmufb$ respectively.

\section{Summary}
\label{s:summary}

The production of $\ttg$ final states 
with a photon with transverse energy greater than 20~\gev\
is observed with a significance of
$\mysignificance\sigma$ in proton--proton collisions at
\mbox{$\sqrt{s}=7~\TeV$} using the ATLAS detector at the CERN LHC. The
dataset used corresponds to an integrated luminosity of
$\runilumi$. The $\ttg$ cross section per lepton flavor, determined
in a fiducial kinematic region within the ATLAS acceptance defined in
Sec.~\ref{s:fiducial}, is measured to be $\xsecexpfb$ in good
agreement with the theoretical prediction.

\section{Acknowledgements}

We thank CERN for the very successful operation of the LHC, as well as the
support staff from our institutions without whom ATLAS could not be
operated efficiently.

We acknowledge the support of ANPCyT, Argentina; YerPhI, Armenia; ARC,
Australia; BMWFW and FWF, Austria; ANAS, Azerbaijan; SSTC, Belarus; CNPq and FAPESP,
Brazil; NSERC, NRC and CFI, Canada; CERN; CONICYT, Chile; CAS, MOST and NSFC,
China; COLCIENCIAS, Colombia; MSMT CR, MPO CR and VSC CR, Czech Republic;
DNRF, DNSRC and Lundbeck Foundation, Denmark; EPLANET, ERC and NSRF, European Union;
IN2P3-CNRS, CEA-DSM/IRFU, France; GNSF, Georgia; BMBF, DFG, HGF, MPG and AvH
Foundation, Germany; GSRT and NSRF, Greece; RGC, Hong Kong SAR, China; ISF, MINERVA, GIF, I-CORE and Benoziyo Center,
Israel; INFN, Italy; MEXT and JSPS, Japan; CNRST, Morocco; FOM and NWO,
Netherlands; BRF and RCN, Norway; MNiSW and NCN, Poland; GRICES and FCT, Portugal; MNE/IFA,
Romania; MES of Russia and NRC KI, Russian Federation; JINR; MSTD,
Serbia; MSSR, Slovakia; ARRS and MIZ\v{S}, Slovenia; DST/NRF, South Africa;
MINECO, Spain; SRC and Wallenberg Foundation, Sweden; SER, SNSF and Cantons of
Bern and Geneva, Switzerland; NSC, Taiwan; TAEK, Turkey; STFC, the Royal
Society and Leverhulme Trust, United Kingdom; DOE and NSF, United States of
America.

The crucial computing support from all WLCG partners is acknowledged
gratefully, in particular from CERN and the ATLAS Tier-1 facilities at
TRIUMF (Canada), NDGF (Denmark, Norway, Sweden), CC-IN2P3 (France),
KIT/GridKA (Germany), INFN-CNAF (Italy), NL-T1 (Netherlands), PIC (Spain),
ASGC (Taiwan), RAL (UK) and BNL (USA) and in the Tier-2 facilities
worldwide.

\bibliographystyle{myamsplain}
\bibliography{main}

\providecommand{\bysame}{\leavevmode\hbox to3em{\hrulefill}\thinspace}
\providecommand{\MR}{\relax\ifhmode\unskip\space\fi MR }
\providecommand{\MRhref}[2]{%
  \href{http://www.ams.org/mathscinet-getitem?mr=#1}{#2}
}
\providecommand{\href}[2]{#2}
\begin{thebibliography}{10}

\bibitem{Lillie:2007hd}
B.~Lillie, J.~Shu, and T.~M.~P. Tait, J. High Energy Phys.~04, 087  (2008),
  \href{http://arxiv.org/abs/0712.3057}{arXiv:0712.3057}.

\bibitem{CDFttg3}
{CDF Collaboration, T. Aaltonen et al.}, Phys. Rev.~D \textbf{84}, 031104
  (2011), \href{http://arxiv.org/abs/1106.3970}{arXiv:1106.3970}.

\bibitem{Aad:2008zzm}
{ATLAS Collaboration}, JINST \textbf{3}, S08003  (2008),
  \href{http://dx.doi.org/10.1088/1748-0221/3/08/S08003}{doi:10.1088/1748-0221/3/08/S08003}.

\bibitem{Aad:2013ucp}
{ATLAS Collaboration}, Eur.~Phys.~J.~C \textbf{73}, 2518  (2013),
  \href{http://arxiv.org/abs/1302.4393}{arXiv:1302.4393}.

\bibitem{:2010wqa}
{ATLAS Collaboration}, Eur.~Phys.~J.~C \textbf{70}, 823  (2010),
  \href{http://arxiv.org/abs/arXiv:1005.4568}{arXiv:1005.4568}.

\bibitem{AGO-0301}
{GEANT4 Collaboration, S. Agostinelli et al.}, Nucl.~Instrum.~Meth.~A
  \textbf{506}, 250  (2003),
  \href{http://dx.doi.org/10.1016/S0168-9002(03)01368-8}{doi:10.1016/S0168-9002(03)01368-8}.

\bibitem{Sjostrand:2006za}
T.~Sj{\"o}strand, S.~Mrenna, and P.~Skands, J.~High~Energy~Phys.~05, 026
  (2006), \href{http://arxiv.org/abs/hep-ph/0603175}{arXiv:0603175 [hep-ph]}.

\bibitem{ATLAS-CONF-2010-031}
{ATLAS Collaboration},
  \href{http://cdsweb.cern.ch/record/1277665}{ATLAS-CONF-2010-031} (2010),
  \href{http://cdsweb.cern.ch/record/1277665}{http://cdsweb.cern.ch/record/1277665}.

\bibitem{Whizard}
W.~Kilian, T.~Ohl, and J.~Reuter, Eur.~Phys.~J.~C \textbf{71}, 1742  (2011),
  \href{http://arxiv.org/abs/0708.4233}{arXiv:0708.4233}.

\bibitem{Omega}
M.~Moretti, T.~Ohl, and J.~Reuter, LC-TOOL-2001-040-rev (2001),
  \href{http://arxiv.org/abs/hep-ph/0102195}{arXiv:0102195 [hep-ph]}.

\bibitem{Maltoni:2002qb}
F.~Maltoni and T.~Stelzer, J.~High~Energy~Phys.~02, 027  (2003),
  \href{http://arxiv.org/abs/hep-ph/0208156}{arXiv:0208156 [hep-ph]}.

\bibitem{cteq6}
{J.~Pumplin et al.}, J.~High~Energy~Phys.~07, 012  (2002),
  \href{http://arxiv.org/abs/hep-ph/0201195}{arXiv:0201195 [hep-ph]}.

\bibitem{COR-0001}
{R.~Corcella et al.}, J.~High~Energy~Phys.~01, 010  (2001),
  \href{http://arxiv.org/abs/hep-ph/0011363}{arXiv:0011363 [hep-ph]}.

\bibitem{JButterworth:1996zw}
J.~M. Butterworth, J.~R. Forshaw, and M.~H. Seymour, Z.~Phys.~C \textbf{72},
  637  (1996), \href{http://arxiv.org/abs/hep-ph/9601371}{arXiv:9601371
  [hep-ph]}.

\bibitem{ATLAS:2011gmi}
{ATLAS Collaboration},
  \href{http://cdsweb.cern.ch/record/1345343}{ATL-PHYS-PUB-2011-008} (2011),
  \href{http://cdsweb.cern.ch/record/1345343}{http://cdsweb.cern.ch/record/1345343}.

\bibitem{Skands:2010ak}
P.~Z. Skands, Phys.~Rev.~D \textbf{82}, 074018  (2010),
  \href{http://arxiv.org/abs/arXiv:1005.3457}{arXiv:1005.3457}.

\bibitem{Photos}
P.~Golonka and Z.~Was, Eur.~Phys.~J.~C \textbf{45}, 97  (2006),
  \href{http://arxiv.org/abs/0506026}{arXiv:0506026 [hep-ph]}.

\bibitem{Melnikov:private}
K.~Melnikov, A.~Scharf, and M.~Schulze, Private communication.

\bibitem{Frixione:2002ik}
S.~Frixione and B.~R. Webber, J.~High~Energy~Phys.~06, 029  (2002),
  \href{http://arxiv.org/abs/hep-ph/0204244}{arXiv:0204244 [hep-ph]}.

\bibitem{Frixione:2003ei}
S.~Frixione, P.~Nason, and B.~R. Webber, J.~High~Energy~Phys.~08, 007  (2003),
  \href{http://arxiv.org/abs/hep-ph/0305252}{arXiv:0305252 [hep-ph]}.

\bibitem{Nadolsky:2008zw}
{P.~M.~Nadolsky et al.}, Phys.~Rev.~D \textbf{78}, 013004  (2008),
  \href{http://arxiv.org/abs/0802.0007}{arXiv:0802.0007}.

\bibitem{Kersevan:2013ji}
B.~P. Kersevan and E.~Richter-Was, {Comput.~Phys.~Commun.} \textbf{184}, 919
  (2013), \href{http://arxiv.org/abs/hep-ph/0405247}{arXiv:0405247 [hep-ph]}.

\bibitem{ATLAS:2012al}
{ATLAS Collaboration}, Eur.~Phys.~J.~C \textbf{72}, 2043  (2012),
  \href{http://arxiv.org/abs/1203.5015}{arXiv:1203.5015}.

\bibitem{ATLAS:2011zja}
{ATLAS Collaboration},
  \href{http://cdsweb.cern.ch/record/1363300}{ATL-PHYS-PUB-2011-009},
  \href{http://cdsweb.cern.ch/record/1363300}{http://cdsweb.cern.ch/record/1363300}.

\bibitem{Cacciari:2011hy}
{M.~Cacciari et al.}, Phys.~Lett.~B \textbf{710}, 612  (2012),
  \href{http://arxiv.org/abs/1111.5869}{arXiv:1111.5869}.

\bibitem{Baernreuther:2012ws}
P.~Baernreuther, M.~Czakon, and A.~Mitov, Phys.~Rev.~Lett. \textbf{109}, 132001
   (2012), \href{http://arxiv.org/abs/1204.5201}{arXiv:1204.5201}.

\bibitem{Czakon:2012zr}
M.~Czakon and A.~Mitov, J.~High~Energy~Phys.~12, 054  (2012),
  \href{http://arxiv.org/abs/1207.0236}{arXiv:1207.0236}.

\bibitem{Czakon:2012pz}
M.~Czakon and A.~Mitov, J.~High~Energy~Phys.~01, 080  (2013),
  \href{http://arxiv.org/abs/1210.6832}{arXiv:1210.6832}.

\bibitem{Czakon:2013goa}
M.~Czakon, P.~Fiedler, and A.~Mitov, Phys.~Rev.~Lett. \textbf{110}, 252004
  (2013), \href{http://arxiv.org/abs/1303.6254}{arXiv:1303.6254}.

\bibitem{Czakon:2011xx}
M.~Czakon and A.~Mitov, Comput.~Phys.~Commun. \textbf{185}, 2930  (2014),
  \href{http://arxiv.org/abs/1112.5675}{arXiv:1112.5675}.

\bibitem{MAN-0301}
{M.~L.~Mangano et al.}, J.~High~Energy~Phys.~07, 001  (2003),
  \href{http://arxiv.org/abs/hep-ph/0206293}{arXiv:0206293 [hep-ph]}.

\bibitem{Martin:1998sq}
A.~D. Martin, R.~G. Roberts, W.~J. Stirling, and R.~S. Thorne, Eur.~Phys.~J.~C
  \textbf{4}, 463  (1998),
  \href{http://arxiv.org/abs/hep-ph/9803445}{arXiv:9803445 [hep-ph]}.

\bibitem{GLE-0901}
{T.~Gleisberg et al.}, J.~High~Energy~Phys.~02, 007  (2009),
  \href{http://arxiv.org/abs/arXiv:0811.4622}{arXiv:0811.4622}.

\bibitem{Lai:2010vv}
{H.~L.~Lai et al.}, Phys.~Rev.~D \textbf{82}, 074024  (2010),
  \href{http://arxiv.org/abs/arXiv:1007.2241}{arXiv:1007.2241}.

\bibitem{Frixione:2008yi}
S.~Frixione, E.~Laenen, P.~Motylinski, B.~R. Webber, and C.D.~White, J.~High~Energy~Phys.~07, 029  (2008),
  \href{http://arxiv.org/abs/0805.3067}{arXiv:0805.3067 [hep-ph]}.

\bibitem{Frixione:2005vw}
S.~Frixione, E.~Laenen, P.~Motylinski, and B.~R. Webber, J.~High~Energy~Phys.~03, 092  (2006),
  \href{http://arxiv.org/abs/hep-ph/0512250}{arXiv:0512250 [hep-ph]}.
  
\bibitem{ATLASTrigPerformance2}
{ATLAS Collaboration}, Eur.~Phys.~J.~C \textbf{72}, 1849  (2012),
  \href{http://arxiv.org/abs/1110.1530}{arXiv:1110.1530}.

\bibitem{ATLAS:2012ar}
{ATLAS Collaboration},
  \href{http://cdsweb.cern.ch/record/1473426}{ATLAS-CONF-2012-123},
  \href{http://cdsweb.cern.ch/record/1473426}{http://cdsweb.cern.ch/record/1473426}.

\bibitem{ATLAS:2012ar2}
{ATLAS Collaboration}, Phys.~Rev.~D \textbf{85}, 092014  (2012),
  \href{http://arxiv.org/abs/1203.3161}{arXiv:1203.3161}.

\bibitem{Aad:2010sp}
{ATLAS Collaboration}, Phys.~Rev.~D \textbf{83}, 052005  (2011),
  \href{http://arxiv.org/abs/1012.4389}{arXiv:1012.4389}.

\bibitem{ATLASElectronPerformance}
{ATLAS Collaboration}, Eur.~Phys.~J.~C \textbf{74}, 2941  (2014),
  \href{http://arxiv.org/abs/arXiv:1404.2240}{arXiv:1404.2240}.

\bibitem{Aad:2014zya}
{ATLAS Collaboration}, Eur.~Phys.~J.~C \textbf{74}, 3034  (2014),
  \href{http://arxiv.org/abs/arXiv:1404.4562}{arXiv:1404.4562}.

\bibitem{Aad:2011he}
{ATLAS Collaboration}, Eur.~Phys.~J.~C \textbf{73}, 2304  (2013),
  \href{http://arxiv.org/abs/arXiv:1112.6426}{arXiv:1112.6426}.

\bibitem{Cojocaru:2004jk}
C.~Cojocaru et~al., Nucl.~Instrum.~Meth.~A \textbf{531}, 481  (2004),
  \href{http://arxiv.org/abs/physics/0407009}{arXiv:physics/0407009}.

\bibitem{Lampl:2008zz}
W.~Lampl et~al.,  (2008),
  \href{http://inspirehep.net/record/807147}{ATL-LARG-PUB-2008-002}.

\bibitem{antikt}
M.~Cacciari, G.~P. Salam, and G.~Soyez, J.~High~Energy~Phys.~04, 063  (2008),
  \href{http://arxiv.org/abs/arXiv:0802.1189}{arXiv:0802.1189}.

\bibitem{Aad:2012qf}
{ATLAS Collaboration}, Phys.~Lett.~B \textbf{711}, 244  (2012),
  \href{http://arxiv.org/abs/arXiv:1201.1889}{arXiv:1201.1889}.

\bibitem{ATLAS-CONF-2012-040}
{ATLAS Collaboration},
  \href{http://cdsweb.cern.ch/record/1435194}{ATLAS-CONF-2012-040},
  \href{http://cdsweb.cern.ch/record/1435194}{http://cdsweb.cern.ch/record/1435194}.

\bibitem{ATLAS-CONF-2012-043}
{ATLAS Collaboration},
  \href{http://cdsweb.cern.ch/record/1435197/}{ATLAS-CONF-2012-043},
  \href{http://cdsweb.cern.ch/record/1435197}{http://cdsweb.cern.ch/record/1435197}.

\bibitem{ATLAS-CONF-2011-089}
{ATLAS Collaboration},
  \href{http://cdsweb.cern.ch/record/1356198/}{ATLAS-CONF-2011-089},
  \href{http://cdsweb.cern.ch/record/1356198/}{http://cdsweb.cern.ch/record/1356198/}.

\bibitem{Aad:2012re}
{ATLAS Collaboration}, Eur.~Phys.~J.~C \textbf{72}, 1844  (2012),
  \href{http://arxiv.org/abs/arXiv:1108.5602}{arXiv:1108.5602}.

\bibitem{Aad:2014bia}
{ATLAS Collaboration}, Eur.~Phys.~J.~C \textbf{75}, 17 (2015),
  \href{http://arxiv.org/abs/arXiv:1406.0076}{arXiv:1406.0076}.

\bibitem{ATLASElectronPerformance2010}
{ATLAS Collaboration}, Eur.~Phys.~J.~C \textbf{72}, 1909  (2012),
  \href{http://arxiv.org/abs/arXiv:1110.3174}{arXiv:1110.3174}.

\bibitem{RooFit}
W.~Verkerke and D.~Kirkby, {eConf} \textbf{C0303241}, MOLT007  (2003),
  \href{http://arxiv.org/abs/physics/0306116}{arXiv:physics/0306116
  [physics.data-an]}.

\bibitem{RooStats}
{L.~Moneta et al.}, PoS ACAT2010 (2010),
  \href{http://arxiv.org/abs/1009.1003}{arXiv:1009.1003 [physics.data-an]}.

\bibitem{feldman-cousins}
G.~J. Feldman and R.~D. Cousins, Phys.~Rev.~D \textbf{57}, 3873  (1998),
  \href{http://arxiv.org/abs/physics/9711021}{arXiv:9711021 [physics.data-an]}.

\bibitem{Oreglia:1980cs}
M.~Oreglia, Ph.D. Thesis, SLAC-236 (1980), Appendix~D., SLAC.

\bibitem{Gaiser:1982yw}
J.~Gaiser, Ph.D. Thesis, SLAC-R-255 (1982), Appendix~F., SLAC.

\bibitem{Aad:2010ey}
{ATLAS Collaboration}, Eur.~Phys.~J.~C \textbf{71}, 1577  (2011),
  \href{http://arxiv.org/abs/1012.1792}{arXiv:1012.1792}.

\bibitem{Berends:1990ax}
F.~A. Berends, H.~Kuijf, B.~Tausk, and W.~T. Giele, Nucl.~Phys.~B \textbf{357},
  32  (1991),
  \href{http://dx.doi.org/10.1016/0550-3213(91)90458-A}{doi:10.1016/0550-3213(91)90458-A}.

\bibitem{Kidonakis:2011wy}
N.~Kidonakis, Phys.~Rev.~D \textbf{83}, 091503  (2011),
  \href{http://arxiv.org/abs/1103.2792}{arXiv:1103.2792}.

\bibitem{Kidonakis:2010tc}
N.~Kidonakis, Phys.~Rev.~D \textbf{81}, 054028  (2010),
  \href{http://arxiv.org/abs/1001.5034}{arXiv:1001.5034}.

\bibitem{Kidonakis:2010ux}
N.~Kidonakis, Phys.~Rev.~D \textbf{82}, 054018  (2010),
  \href{http://arxiv.org/abs/1005.4451}{arXiv:1005.4451}.

\bibitem{Anastasiou:2003ds}
C.~Anastasiou, L.~J. Dixon, F.~Melnikov, and K.~Petriello, Phys.~Rev.~D
  \textbf{69}, 094008  (2004),
  \href{http://arxiv.org/abs/hep-ph/0312266}{arXiv:0312266 [hep-ph]}.

\bibitem{Campbell:2011bn}
J.~M. Campbell, R.~K. Ellis, and C.~Williams, J.~High~Energy~Phys.~07, 018
  (2011), \href{http://arxiv.org/abs/1105.0020}{arXiv:1105.0020}.

\bibitem{Botje:2011sn}
M.~Botje et~al., \href{http://arxiv.org/abs/1101.0538}{arXiv:1101.0538}.

\bibitem{Martin:2009iq}
A.~D. Martin, W.~J. Stirling, R.~S. Thorne, and G.~Watt, Eur.~Phys.~J.~C
  \textbf{63}, 189  (2009),
  \href{http://arxiv.org/abs/arXiv:0901.0002}{arXiv:0901.0002}.

\bibitem{Martin:2009bu}
A.~D. Martin, W.~J. Stirling, R.~S. Thorne, and G.~Watt, {Eur.~Phys.~J.~C}
  \textbf{64}, 653  (2009),
  \href{http://arxiv.org/abs/0905.3531}{arXiv:0905.3531}.

\bibitem{Gao:2013xoa}
{J.~Gao et al.}, Phys.~Rev.~D \textbf{89}, 033009  (2014),
  \href{http://arxiv.org/abs/1302.6246}{arXiv:1302.6246}.

\bibitem{Ball:2012cx}
R.~D. Ball et~al., Nucl.~Phys.~B \textbf{867}, 244  (2013),
  \href{http://arxiv.org/abs/1207.1303}{arXiv:1207.1303}.

\bibitem{Aad:2012ag}
{ATLAS Collaboration}, Eur.~Phys.~J.~C \textbf{73}, 2306  (2013),
  \href{http://arxiv.org/abs/arXiv:1210.6210}{arXiv:1210.6210}.

\bibitem{Wilks:1938dza}
S.~S. Wilks, Annals~Math.~Statist. \textbf{9}, 60  (1938), ~1,
  \href{http://dx.doi.org/10.1214/aoms/1177732360}{doi:10.1214/aoms/1177732360}.

\bibitem{Melnikov}
K.~Melnikov, A.~Scharf, and M.~Schulze, Phys.~Rev.~D \textbf{83}, 074013
  (2011), \href{http://arxiv.org/abs/1102.1967}{arXiv:1102.1967}.

\bibitem{FRI-pQCD}
S.~Frixione, Phys.~Lett.~B \textbf{429}, 369  (1998),
  \href{http://arxiv.org/abs/hep-ph/9801442}{arXiv:9801442 [hep-ph]}.

\end{thebibliography}

\clearpage
\appendix
\part*{Appendices}

\section{$\ttg$ Monte Carlo samples}
\label{s:ttgamma}

Signal \mbox{$\ttg$ events} with single-lepton ($\ell\nu_\ell
q\bar{q'}b\bar{b}\gamma$, $\ell\equiv$~$e,\,\mu,\,\tau$) or dilepton
($\ell\nu_\ell \ell'\nu_{\ell'}b\bar{b}\gamma$, $\ell/\ell'\equiv$~$e,\,\mu,\,\tau$)
final states are simulated with two independent
leading-order (LO) matrix element~(ME) Monte Carlo
generators, \mbox{\whizardv~\cite{Whizard,Omega}} and
  \mbox{\madgraphv~\cite{Maltoni:2002qb}}, both using the
  CTEQ6L1~\cite{cteq6} LO parton distribution function (PDF)
set. Both calculations take into account interference effects between
radiative top-quark production and decay processes.

\subsection{Leading-order calculations: \whizard\ and \madgraph}
\label{s:ttgammalo}

In the \whizard\ $\ttg$ sample, the minimum transverse momentum of all
outgoing partons except for the photon is set to $10~\GeV$.  The
transverse momentum of the photon is required to be larger than
$8~\GeV$. The invariant mass of the photon and any charged particle
($u$-, $d$-, $c$- and $s$-quarks, electrons, muons, and $\tau$
leptons) is required to be larger than $5~\GeV$. To avoid infrared and
collinear divergences, the following invariant masses are also
required to be larger than $5~\GeV$: $m(q_1, q_2)$, $m(g_1, q_1)$,
$m(g_1, q_2)$, $m(g_2, q_1)$, and $m(g_2, q_2)$, where $q_1$ and $q_2$
are the quarks from the hadronic decay of one $W$ boson, and $g_1$ and
$g_2$ are the gluons initiating the $gg\rightarrow \ttg$ process. 
For each incoming quark $Q_i$ (\mbox{$u$-}, \mbox{$d$-}, \mbox{$c$-}, \mbox{$s$-} and \mbox{$b$-quark}),
the invariant mass $m(Q_i, q_j)$ is required to be larger than
$5~\GeV$ if $q_j$ is the same type of parton as $Q_i$. 
The renormalization scale is set to 2$m_t$, and
the factorization scale is set to the partonic center-of-mass energy
$\sqrt{\hat s}$. The cross section is 648~fb when summing over all
three lepton flavors for the single-lepton ($e$, $\mu$, $\tau$) and
188~fb for the dilepton $\ttg$ final states.

In the \madgraph\ $\ttg$ sample, the minimum transverse momentum is set
to $15~\GeV$ for $u$-, $d$-, $c$- and $s$-quarks, as well as for photons,
electrons, muons and $\tau$ leptons. The distance in 
$\eta$--$\phi$ space 
between all these
particles is required to be $\deltarmadgraph > 0.2$. 
For $b$-quarks, no requirement is placed on the transverse momentum or on the pseudorapidity. Leptons and
photons are required to have $|\eta|<2.8$, while \mbox{$u$-,} $d$-, $c$- and
$s$-quarks are required to have $|\eta|<5.0$. The renormalization and
factorization scales are set to $m_t$. 
The cross section is 445~fb when summing over all three lepton
flavors for the single-lepton and 131~fb for the dilepton $\ttg$
final states. 

\subsection{Next-to-leading-order calculation}
\label{s:ttgammanlo}

The NLO QCD calculation of top-quark pair production in association
with a hard photon is described in Ref.~\cite{Melnikov} for
$\sqrt{s}=14~\TeV$. A dedicated calculation at $\sqrt{s}=7~\TeV$ both at LO and at NLO has
been performed for this analysis~\cite{Melnikov:private} 
for the $pp\rightarrow
b\mu^+\nu_\mu\bar{b}jj\gamma$ channel 
using the same settings for the renormalization and factorization
scale as in the \whizard\ $\ttg$ calculation.

The following NLO input parameters are used: top-quark mass $m_t =
172~\GeV$, top-quark width $\Gamma_t = 1.3237~\GeV$, $W$-boson mass
$m_W = 80.419~\GeV$, $W$-boson width $\Gamma_W = 2.14~\GeV$,
fine-structure constant $\alpha = 1/137$. The strong-coupling constant
$\alpha_{\rm s}(\mu)$ is evaluated using the two-loop running from
$\alpha_{\rm s}(m_Z)$ as specified in the MSTW2008 NLO PDF. 
Jets are defined using the anti-$k_t$ algorithm with a distance parameter $R~=~0.4$. 
The photon is required to be separated from hadronic activity as
defined in Ref.~\cite{FRI-pQCD}. 

The phase-space requirements used in the $\sqrt{s}=7~\TeV$ theory LO
and NLO calculations are described below. The muon is required to have
$\pT(\mu) > 20~\GeV$ and $|\eta(\mu)| < 2.5$. The missing transverse
momentum is required to be $\met > 25~\GeV$ and $\met+m_{\rm T}^W >
60~\GeV$, where $m_{\rm T}^W$ is the $W$-boson transverse mass. Jets
are required to have $\pT(j) > 25~\GeV$ and $|\eta(j)| < 2.5$. The
photon is required to have $\pT(\gamma) > 15~\GeV$ and $|\eta(\gamma)|
< 1.37$ or $1.52 < |\eta(\gamma)| < 2.37$. The objects are required to
be separated in $\Delta R$: $\dr({\rm jets}) > 0.4$, $\dr(\mu,{\rm
  jets}) > 0.4$, $\dr(\gamma,\mu) > 0.4$, $\dr(\gamma,{\rm
  jets})>0.5$. The event is required to have $N_{\rm jets} \ge 4$.

With the above setup and assuming 100\% efficiencies, $\sigma^{\rm
  NLO}_{\ttg} = 24.5\pmasym{5.6}{4.5}$~pb and $\sigma^{\rm LO}_{\ttg}
= 14.7\pmasym{5.8}{3.8}$~pb. Upper and lower values correspond to
scale variations by a factor of two around $\mu=m_t$. Therefore, for $
\mu=m_t $ the NLO/LO $K$-factor is 1.67. Similarly, for the
\whizard\ Monte Carlo sample scales and NLO calculation at the scale
of $ \mu=m_t $, the NLO/LO $K$-factor is 2.53.

The LO cross sections calculated with the \whizard\ and
\madgraph\ Monte Carlo generators are multiplied by the
corresponding $K$-factors in order to compare with the experimental
measurement.

\onecolumngrid
\clearpage 
\begin{flushleft}
{\Large The ATLAS Collaboration}

\bigskip

G.~Aad$^{\rm 84}$,
B.~Abbott$^{\rm 112}$,
J.~Abdallah$^{\rm 152}$,
S.~Abdel~Khalek$^{\rm 116}$,
O.~Abdinov$^{\rm 11}$,
R.~Aben$^{\rm 106}$,
B.~Abi$^{\rm 113}$,
M.~Abolins$^{\rm 89}$,
O.S.~AbouZeid$^{\rm 159}$,
H.~Abramowicz$^{\rm 154}$
% ...
\bigskip
\\
% ...
$^{5}$ LAPP, CNRS/IN2P3 and Universit{\'e} de Savoie, Annecy-le-Vieux, France\\
$^{6}$ High Energy Physics Division, Argonne National Laboratory, Argonne IL, United States of America\\
% ...
$^{8}$ Department of Physics, The University of Texas at Arlington, Arlington TX, United States of America\\
% ...
$^{11}$ Institute of Physics, Azerbaijan Academy of Sciences, Baku, Azerbaijan\\
% ...
$^{84}$ CPPM, Aix-Marseille Universit{\'e} and CNRS/IN2P3, Marseille, France\\
% ...
$^{89}$ Department of Physics and Astronomy, Michigan State University, East Lansing MI, United States of America\\
% ...
$^{106}$ Nikhef National Institute for Subatomic Physics and University of Amsterdam, Amsterdam, Netherlands\\
% ...
$^{112}$ Homer L. Dodge Department of Physics and Astronomy, University of Oklahoma, Norman OK, United States of America\\
$^{113}$ Department of Physics, Oklahoma State University, Stillwater OK, United States of America\\
% ...
$^{116}$ LAL, Universit{\'e} Paris-Sud and CNRS/IN2P3, Orsay, France\\
% ...
$^{152}$ Institute of Physics, Academia Sinica, Taipei, Taiwan\\
% ...
$^{154}$ Raymond and Beverly Sackler School of Physics and Astronomy, Tel Aviv University, Tel Aviv, Israel\\
% ...
$^{159}$ Department of Physics, University of Toronto, Toronto ON, Canada\\
% ...
$^{s}$ Also at CERN, Geneva, Switzerland\\
% ...
\end{flushleft}

\end{document}